\documentclass[11pt, letterpaper]{article}
\usepackage[utf8]{inputenc}
\usepackage[margin=1.5cm]{geometry}
\usepackage{titlesec}
\usepackage{tabu}
\usepackage{enumitem}
\usepackage{amssymb}
\usepackage{xcolor}
\newlist{selectlist}{itemize}{2}
\setlist[selectlist]{label=$\square$,leftmargin=*,noitemsep,topsep=0pt}

\usepackage{lmodern}

\usepackage{hyperref}
\hypersetup{
    colorlinks=true,
    linkcolor=blue,
    filecolor=magenta,      
    urlcolor=blue,
}

\usepackage[AMcore,AMbiblio,AMref,AMmath,AMtikz,optEnglish,optBiber]{AMlatex}

\usepgfplotslibrary{groupplots}
\tikzset{external/system call={pdflatex \tikzexternalcheckshellescape -halt-on-error
		-interaction=batchmode -jobname "\image" "\texsource" --extra-mem-bot=100000000 --extra-mem-top=100000000 --pool-size=10000000 --buf-size=10000000}}

\makeatletter
\let\blx@rerun@biber\relax
\makeatother

\usepackage{graphicx}

\graphicspath{{figures/},{figures/hw_description},{figures/fw_description},{figures/assembly},{figures/validation},{figures/operating_instructions}}
 
\usepackage{float}

\usepackage{colortbl}
\usepackage{multirow}

\usepackage{tcolorbox}
\newtcolorbox{NOTES}{parbox=false,
	boxrule=0pt,leftrule=3mm,rightrule=3mm,boxsep=0pt,arc=0pt,outer arc=0pt,
	left=3mm,right=3mm,top=3mm,bottom=3mm,toptitle=1mm,bottomtitle=1mm,oversize,
	colback=blue!5!white,colframe=blue}

\titleformat{\section}[block]{\hspace{1em}\bfseries}{\thesection.}{0.5em}{} 
\titleformat{\subsection}[block]{\hspace{1em}}{\thesubsection}{0.5em}{}

\def\ucr{\scalebox{0.8}{\stackinset{c}{}{c}{-.2pt}{%
			\textcolor{white}{\sffamily\bfseries\scriptsize ?}}{%
			\rotatebox{45}{$\blacksquare$}}}}

\usepackage{listings}
\usepackage{stackengine}
\definecolor{codegray}{rgb}{0.5,0.5,0.5}
\definecolor{OASISUROS}{rgb}{0.0274,0.3804,0.6196}
\newtcolorbox
{OASISPrompt}[2][]
{title=\textcolor{white}{#2},
	width=0.9\textwidth,
	left=0.025\textwidth,
	right=0.025\textwidth,
	fonttitle=\bfseries,
	coltitle=black,
	colframe=OASISUROS!85!white,
	colback=gray!5!white,
	before=\noindent\centering
} 
\newtcolorbox
{fileContents}[2][]
{title=\texttt{#2},
	width=0.9\textwidth,
	left=0.025\textwidth,
	right=0.025\textwidth,
	fonttitle=\bfseries,
	coltitle=black,
	colframe=TUMBlue1!25!white,
	colback=gray!5!white,
	before=\noindent\centering
} 

\lstdefinestyle{OASISPrompt}{
	basicstyle=\fontfamily{pcr}\footnotesize,
	breakatwhitespace=false,         
	breaklines=true,                 
	captionpos=b,                    
	keepspaces=true,                                 
	showspaces=false,                
	showstringspaces=false,
	showtabs=false,                  
	tabsize=2,
	inputencoding=latin1
}

\lstdefinestyle{fileContent}{
	numberstyle=\tiny\color{codegray},
	basicstyle=\fontfamily{pcr}\footnotesize,
	breakatwhitespace=false,         
	breaklines=true,                 
	captionpos=b,                    
	keepspaces=true,                 
	numbers=left,                    
	numbersep=5pt,                  
	showspaces=false,                
	showstringspaces=false,
	showtabs=false,                  
	tabsize=2
}

\newcommand{\SpecTable}{\bgroup\def\arraystretch{1.15}\setlength\tabcolsep{2mm}\begin{tabular}{|l|l|}
		\hline Available channels & 8 IEPE channels with BNC connector \\\hline
		Sampling frequency @ 8 ch. & \SI{36}{\kilo\hertz} per channel @ 18-bit resolution \\\hline
		Voltage ranges & $\pm\SI{2.5}{\volt}$, $\pm\SI{5}{\volt}$, $\pm\SI{6.25}{\volt}$, $\pm\SI{10}{\volt}$, $\pm\SI{12.5}{\volt}$ \\\hline
		Bill of materials cost & $\approx$ \textdollar\,220 / 200\,\texteuro \\\hline
\end{tabular}\egroup}

\begin{document}

\setlength{\parindent}{0pt}
\setlength{\parskip}{10pt}

% !TeX spellcheck = en_US

\begin{flushleft}
%Insert title
\textbf{Article title}\\ OASIS-UROS: Open Acquisition System for IEPE Sensors - Upgraded, Refined, and Overhauled Software

%Insert Authors
\textbf{Authors}\\ Oliver Maximilian Zobel$^1$, Johannes Maierhofer$^2$, Andreas Köstler$^1$, Daniel J. Rixen$^1$

%Insert Affiliations
\textbf{Affiliations}\\ $^1$: Chair of Applied Mechanics, TUM School of Engineering and Design,
Technical University of Munich, Boltzmannstr. 15, 85748 Garching, Germany \\
$^2$: Maierhofer-Technology, Stifterweg 9, 94474 Vilshofen

%Insert Contact Email
\textbf{Corresponding author’s email address}\\ oliver.zobel@tum.de

%Insert Abstract
\textbf{Abstract}\\ 
\textit{OASIS-UROS} continues the previously published \textit{Open Acquisition System for IEPE Sensors (OASIS)}. While still building on the \textit{ESP32} microcontroller, this version improves the overall performance by switching to an SD card caching system and upgrading the analog-digital converter to an \textit{AD7606C-18}, which has a higher resolution, provides eight channels, oversampling, and software-adjustable voltage ranges. Also improved is the IEPE front-end and power supply, as well as the firmware of the acquisition system, which can now achieve a sample rate of up to \SI{36}{\kilo\hertz} while sampling all eight channels. This paper documents the hardware and software of \textit{OASIS-UROS} and provides all materials required to reproduce the open acquisition system. Lastly, the system was validated against commercial hardware and software in an experimental modal analysis context. This showed that the system performs close to the commercial one in some aspects with respect to the utilized test case. While \textit{OASIS-UROS} cannot match the full performance of the commercial system, the developed system can be a viable alternative for students, people in academia, or smaller companies that have a constrained budget or require complete insight as well as adaptability of the hardware and software.

%Insert Keywords
\textbf{Keywords}\\ Open-Source, Acquisition Hardware, Data Acquisition, Measurement Equipment, Experimental Dynamics, Vibration Analysis

\textbf{Specifications table}\\
\vskip 0.2cm
\tabulinesep=1ex
\begin{tabu} to \linewidth {|X|X[3,l]|}
	\hline \textbf{Hardware name} & OASIS-UROS \\
	\hline \textbf{Subject area} & Educational tools and open source alternatives to existing \\
	\hline \textbf{Hardware type} & Measuring physical properties and in-lab sensors \\ 
	\hline \textbf{Closest commercial analog} &
	OROS MODS OR10-DAQ-8 or similar compact acquisition systems with 8 IEPE channels and compact form factor\\
	\hline \textbf{Open source license} & CC-BY 4.0 (Hardware) / MIT (Software)\\
	\hline \textbf{Cost of hardware} & $\approx$ \textdollar\,220 / 200\,\texteuro \\
	\hline \textbf{Source file repository} & At time of publication: \url{https://doi.org/10.5281/zenodo.13763227}~\cite{OASIS-UROS-V1.1-Zenodo} \newline
	Most current version: \url{https://gitlab.com/oasis-acquisition}\\
	\hline \textbf{OSHWA certification UID} & DE000150 \\\hline
\end{tabu}
\end{flushleft}

\newpage

% !TeX spellcheck = en_US

\section{Hardware in context}
Driven by advancements in structural health monitoring, predictive maintenance, and industrial automation, the demand for data acquisition systems for vibrational sensor signals is steadily increasing with extremely different requirements. 
The \textit{OASIS-UROS} data acquisition system is built around an \textit{ESP32-S3} microcontroller and an \textit{AD7606C-18} ADC, optimized for capturing high-fidelity vibration data from IEPE (Integrated Electronics Piezo-Electric) sensors. This design reflects a trend in open-source hardware platforms that combine affordability, flexibility, and accessibility for specialized data acquisition tasks. \textit{OASIS} provides users with the ability to record multi-channel, high-resolution analog data, making it suitable for vibration analysis applications, scientific research, and teaching.

It is important to properly position the system presented in this article in terms of its capabilities and limitations within the field of existing systems. A distinction is made between proprietary and open-source solutions. We see the system as somewhat of a bridge between expensive high-end systems and very simple DIY solutions.

\textit{OASIS} is a fraction of the proprietary systems' cost, making it attractive to users with more budget constraints or those looking for a lightweight, customizable system for smaller-scale applications. 

\subsection{Comparison to Proprietary Systems}
Comparing \textit{OASIS-UROS} to established proprietary systems reveals significant differences in capabilities, scalability, and user experience. Many companies offer well-designed hardware. The big differences unveil in their software capabilities.

\noindent\textbf{National Instruments - CompactDAQ}\hspace{0.2cm} The \textit{CompactDAQ} (\textit{cDAQ}) platform from \textit{National Instruments} supports a range of sensor modules, including IEPE inputs, with resolutions up to 24-bit. It is highly modular, with support for various I/O configurations and scalable across multiple modules. The system integrates with National Instruments' \textit{LabVIEW} for data acquisition and processing, but its reliance on \textit{LabVIEW} limits cross-platform flexibility. \textit{cDAQ} offers Python APIs for external access but remains tethered to Windows for full functionality. Hardware costs for IEPE channels are approximately 500 USD per channel.

\noindent\textbf{Brüel \& Kjær - LAN-XI}\hspace{0.2cm} \textit{Brüel \& Kjær's LAN-XI} system is built for precision data capture with support for multiple IEPE channels, offering configurable input ranges and digital filters to improve signal integrity. \textit{LAN-XI} modules provide up to 24-bit resolution and are optimized for environments requiring robustness, such as field testing. However, \textit{LAN-XI} is proprietary, limiting user-level customization and requiring a significant upfront investment for hardware and software.

\noindent\textbf{Siemens LMS - SCADAS}\hspace{0.2cm} \textit{Siemens LMS SCADAS} systems provide high channel counts with 24-bit resolution and sampling rates up to 500 kSps, suitable for large-scale vibration testing. \textit{SCADAS} hardware supports synchronized data acquisition across multiple sensor types, with real-time processing capabilities. A modular, extensive software package is available which offers direct support for Experimental Modal Analysis (EMA), Operational Modal Analysis (OMA), or Transfer Path Analysis (TPA).

\noindent\textbf{Müller BBM - PAK}\hspace{0.2cm} The \textit{PAK} system offers high-resolution data capture with flexible channel configurations. \textit{PAK} supports real-time signal processing and is used primarily for acoustic and vibration testing. 

\noindent\textbf{HBM QuantumX}\hspace{0.2cm} \textit{HBM's QuantumX} is a modular system for data acquisition with support for high precision measurement across different input types, including IEPE. The system features 24-bit resolution and sampling rates up to 100 kSps. \textit{QuantumX} modules are well-suited for high-end industrial and experimental setups. Its strength lies in precise synchronization across multiple modules and high channel scalability.

\newpage

\noindent\textbf{Dewesoft - Sirius Mini}\hspace{0.2cm} The \textit{Sirius Mini} is a compact data acquisition unit offering up to 24-bit resolution with configurable input channels. \textit{Dewesoft's} system is designed for portable applications but retains high precision through its use of \textit{DualCoreADC} technology. Despite its compact form factor, \textit{Sirius Mini} remains a high-cost solution, focused more on portability.

\noindent\textbf{OROS}\hspace{0.2cm} A French company offering systems optimized for dynamic signal analysis in the field of vibrations. \textit{OROS} hardware is mobile, rugged and supports a range of input types, including IEPE. \textit{OROS} provides application oriented software packages for EMA and OMA. The price range is in the high-end region.

\noindent\textbf{Goldammer GmbH}\hspace{0.2cm} \textit{Goldammer GmbH} offers a compact IEPE measurement system designed for precision data acquisition in industrial and laboratory environments. The system supports multiple IEPE channels with 24-bit resolution and sampling rates up to 200 kSps. The system is designed for use with API access for integration with external platforms like MATLAB and Python. Integration is provided for \textit{DaisyLab} and \textit{LabView}. Though not open-source, the hardware is built for reliability and accuracy in demanding applications, while providing flexibility for custom software development through its API. Pricing is competitive within the mid-range market, positioning it as a more affordable alternative to higher-end proprietary systems.

\subsection{Comparison to open-source solutions}
To the authors' knowledge there is no complete open source (hardware and software) IEPE data acquisition system available. 
A few interesting products are listed below. However, they all aim at different use-cases.

\noindent\textbf{MCC 172 DAQ Hat for Raspberry Pi}\hspace{0.2cm} The \textit{MCC 172 DAQ Hat} provides two IEPE channels with 24-bit resolution and sampling rates up to 200 kSps, designed for integration with the \textit{Raspberry Pi}. While the driver software is open-source, the hardware remains proprietary, limiting further customization. Its compact form factor and low cost make it a viable option for small-scale projects, though its performance is constrained by the limitations of the \textit{Raspberry Pi} platform. The \textit{DAQ-Hat} is available for 500 USD without the Raspberry Pi. 

\noindent\textbf{Analog Devices IEPE Data Acquisition Reference Design}\hspace{0.2cm} \textit{Analog Devices} offers a reference design for a 4-channel, 24-bit, 256 kSps IEPE data acquisition system, featuring the \textit{AD7768-4} chip. The design is high-end and provides open schematics but lacks an open-source firmware, limiting its flexibility for custom implementations. The system provides high accuracy and precision but is more suited as a starting point for custom system development than as a ready-to-use platform. A development kit is available for 900 USD. This system is the closest to the \textit{OASIS-UROS} design.

\noindent\textbf{OpenDAQ}\hspace{0.2cm} The \textit{OpenDAQ} is an early days (2008) data acquisition project with a PC/104 (ISA) interface, 16 DIO (digital I/O) ports, and 8 ADC (Analog-to-Digital Converter) inputs. Unfortunately, the last update seemed to be available from 2008. 

\subsection{Conclusion}
In conclusion, the market offers a variety of well-designed hardware solutions for data acquisition, but the critical differentiator among them lies in the software capabilities they provide. Over the past few years, open-source software for structural dynamics has gained significant momentum, exemplified by projects like \textit{pyFBS}~\cite{pyFBS}, which empower users with flexible and customizable tools for analysis. This trend underscores the importance of open-source hardware, as it presents an opportunity to further strengthen the community and enhance the dynamics of open-source structural analysis tools. Now is an opportune moment to invest in open-source hardware development, as it can drive innovation, facilitate collaboration, and provide researchers and practitioners with the accessible resources needed to advance the field of structural health monitoring and analysis.

% !TeX spellcheck = en_US

\newpage

\section{Hardware description}
\textit{OASIS-UROS} is based around the \textit{Analog Devices AD7606C-18} analog-digital-converter (ADC), providing a high quality analog-digital conversion, and the \textit{Espressif ESP32-S3} micro-controller, controlling the data flow. In contrast to traditional DAQ-systems, no FIFO-buffer is used to save costs, space and complexity. A main feature of the board is the on-board IEPE power supply using the \textit{MC34063} series boost-converter and \textit{LT3092} constant current sources. This allows the board to solely operate using a 5V USB power supply.

\begin{figure}[H]
	\centering
	\def\svgwidth{\textwidth}
	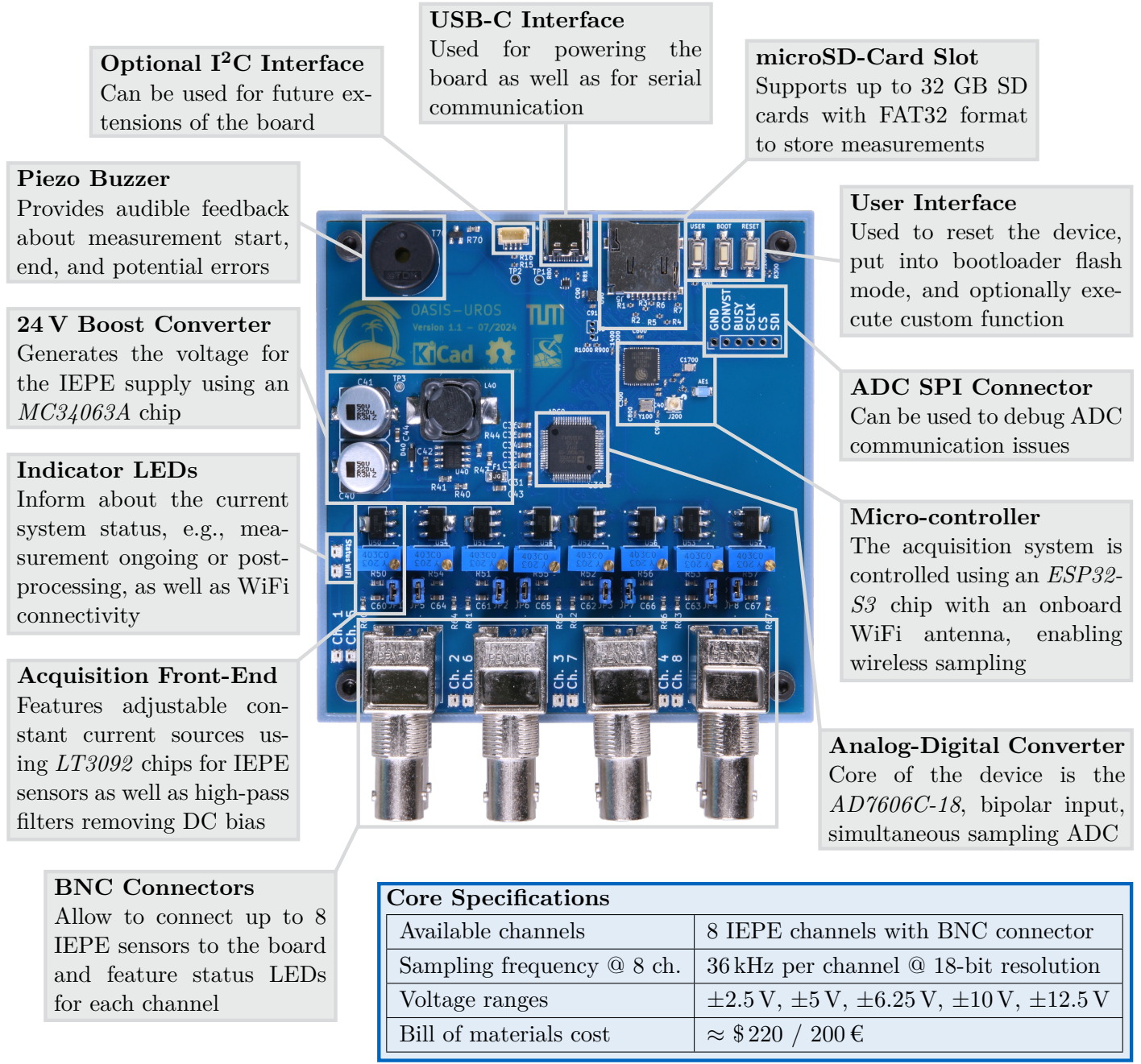\vspace{-0.4cm}
	\caption{Overview of the \textit{OASIS-UROS} board}
	\label{fig:HWOverview}
\end{figure}

A detailed scheme of the 4-layer PCB with the key hardware components is depicted in \cref{fig:HWOverview}.

\newpage
The \textit{AD7606C-18} provides 8 simultaneously sampled channels with individually selectable input ranges. Its integrated analog front end with programmable over-sampling significantly reduces external component count and noise. Each channel is equipped with an acquisition front-end on the \textit{OASIS-UROS} board consisting of a constant current source build using \textit{LT3092} chips. The IEPE source current can be tuned individually for each channel using a potentiometer. All input signals pass a fixed, passive high-pass RC-filter with a \SI{-3}{\decibel} cutoff frequency of \SI{0,8}{\hertz}. This means, \textit{OASIS} always operates in AC mode. To avoid hidden costs, \textit{OASIS} provides BNC connectors directly on the board, i.e., no adapters are needed, and standard sensor equipment can directly be used. 

The interface between ADC and micro-controller is designed such that the ADC is read in data parallel mode, i.e., 8 data lines are connected to one GPIO register of the microcontroller, refered to as Octal SPI. This means, during one read clock cycle, one bit of all 8 channels is read simultaneously. As each channel provides 18-bit resolution, 18 read cycles are necessary to capture all data. Details about the software implementation are described in the next section.

The \textit{ESP32-S3} includes a \textit{RISC-V} co-processor, running at \SI{17.5}{\mega\hertz} which also has access to the GPIO registers. This would allow the outsourcing of the data transfer into the RAM utilizing the co-processor while the main-processor can still be available for user-communication. The authors, however, did not test this approach. Similarly, the \textit{ESP32-S3} includes WiFi connectivity that could be used to acquire sample data wirelessly, as was possible with the original \textit{OASIS} version~\cite{OASISPaper}. Since major parts of the original firmware were rewritten and adjusted, wireless sampling is not yet available with \textit{OASIS-UROS}.
% !TeX spellcheck = en_US
\section{Firmware description}

The firmware for the \textit{OASIS} systems is written in C++ using the Arduino IDE. This code runs on the \textit{ESP32-S3} microcontroller and is mainly responsible for controlling the \textit{AD7606C-18} ADC as well as retrieving the sampled voltage values. The firmware provides an abstraction layer for the user, allowing the hardware to be controlled using high-level commands. Besides requesting sampling, this might be, for instance, configuring the voltage ranges using \texttt{OASIS.SetVoltageRange()}. All available commands are documented in the \textit{OASIS Command Reference}, found in \cref{sec:CommandRef}. While it is possible to control the system manually using the commands provided, using the \textit{OASIS-GUI}, a graphical user interface written in Python, is recommended. Details can be found in operating instructions in \cref{sec:OperatingInstructions}.

When the system is powered on, it reads the device information stored in the EEPROM (refer to the command reference in \cref{sec:CommandRef}) and displays this information via the Serial interface. This information is also used for the \textit{OASIS-GUI} to properly handle the sample conversion process described later. Setting this information is described in \cref{sec:OperatingInstructions}. Besides this, feedback is provided about device operations and whether they were successful. For example, when no SD card is detected, this will be displayed in the startup messages.

\tikzset{external/export next=false}
\begin{OASISPrompt}{OASIS Serial Interface - Startup Sequence}
	\lstset{style=OASISPrompt}\vspace{-0.2cm}\begin{lstlisting}[]
[OASIS] Open Acquisition System for IEPE Sensors starting up...

--------------------------------------------------------------------------
                          OASIS Device Information  
--------------------------------------------------------------------------
[OASIS] Device Architecture is OASIS-UROS (ID 1)
[OASIS] Device Hardware Version is 1.1
[OASIS] Device Firmware Version is 2.0
[OASIS] ADC resolution: 18
[OASIS] TEDS module installed (1=yes, 0=no): 0
[OASIS] WSS module installed (1=yes, 0=no): 0
[OASIS] Device name: OASIS-UROS Alpha
[OASIS] Current device configuration: MUTE: 1, WIFI_EN: 1
--------------------------------------------------------------------------

[OASIS] Resetting ADC...

[OASIS] Setting ADC data out format to 8 lines...
[OASIS] Successfully written 0x18 to address 0x2

[OASIS] Setting voltage range to +/- 2.5V for all channels...
[OASIS] Setting voltage ranges in ADC...
[OASIS] Successfully written 0x0 to address 0x3
[OASIS] Successfully written 0x0 to address 0x4
[OASIS] Successfully written 0x0 to address 0x5
[OASIS] Successfully written 0x0 to address 0x6
[OASIS] Voltage ranges set.

[OASIS] Setting oversampling factor x4 in ADC...
[OASIS] Successfully written 0x2 to address 0x8
[OASIS] Oversampling factor set.

[OASIS] Initializing SD Card...
[OASIS] Detected SD card with 60350MB total space.
[OASIS] Used space: 241MB /29968MB partitioned space.
[OASIS] Free space: 29726MB.

[OASIS] Finished booting.\end{lstlisting}\vspace{-0.2cm}
\end{OASISPrompt}

In the previous version of the \textit{OASIS} firmware, used for the original \textit{OASIS} board~\cite{OASISPaper}, commands can be sent to the system either with a wired connection using the USB-C port and the Serial interface or wirelessly using WiFi and UDP packets. Due to the major rewrite of the firmware, this functionality is not yet available for the \textit{OASIS-UROS} firmware but is planned for future releases.

Reading the sampled values from the ADC in real time and storing them for later use is the most time critical step of data acquisition. The less computation time is required, the faster data can be sampled. Ideally, all samples retrieved from the ADC would just be kept in random access memory (RAM) and processed afterward. However, this is not possible due to the limited RAM of the \textit{ESP32-S3}, which is only \SI{512}{\kilo\byte}. For eight channels with 18-bit data per sample each, only around 28400 samples can be stored in total, assuming the whole RAM could be used.

\begin{figure}[H]
	\centering
	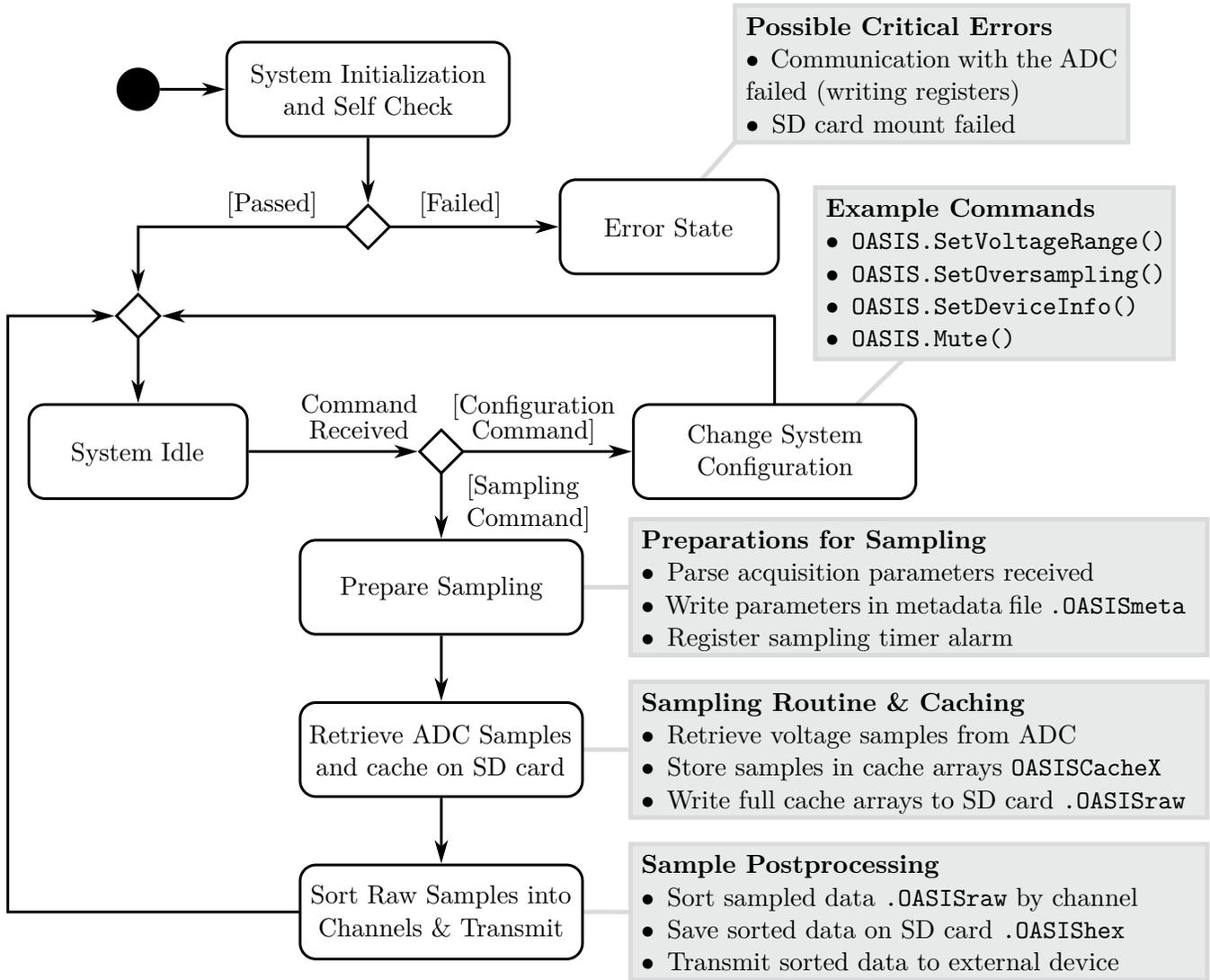
	\caption{State diagram of the \textit{OASIS} firmware depicting the standard sample procedure}
	\label{fig:state_diagram}
\end{figure}

For this reason, the acquired data is cached on an SD card during the data acquisition, effectively allowing for unlimited sampling time, given that microSD cards with large storage capacity are readily available. The sampled data is cached on the microSD card in a raw and unsorted format and post-processed afterward. This is in contrast to the previous \textit{OASIS} version detailed in~\cite{OASISPaper}, where the samples were sent out during the sampling procedure. For convenience, the sampled data can still be transmitted via the Serial interface after the sampling process is completed, e.g., to check signal levels or double impacts. An overview of the fundamental routines is given in the state diagram in \cref{fig:state_diagram}.

If the system is controlled manually, the user should note that voltage ranges and oversampling values are not set using the \texttt{OASIS.Sample()} command, but have to be configured before with the appropriate commands. During the system initialization, all voltage ranges are set to $\pm\SI{2.5}{\volt}$ and the oversampling factor to x4. Further, the ADC is configured to use eight data out lines to send the sampled data to the microcontroller.

Of interest to all users are the files created by \textit{OASIS} on the SD card. When invoking the \texttt{OASIS.Sample()} command, a file name has to be specified, e.g., \texttt{TestFile}. All files belonging to this sample will have this name. The first file created is a metadata file (\texttt{.OASISmeta}) containing all relevant acquisition configurations, e.g., sampling frequency, voltage ranges, device name. Next is the file containing the unprocessed bits of the sample (\texttt{.OASISraw}). This file is usually not required anymore after the sample has been processed by \textit{OASIS}. The sorted data is saved in a file ending with \texttt{.OASIShex} containing the sampled bits in the raw format expected by the \textit{OASIS-GUI}. An example usage is given below, where the user enters their commands as follows:

\tikzset{external/export next=false}
\begin{OASISPrompt}{OASIS Serial Interface - Example Commands}
	\lstset{style=OASISPrompt}\vspace{-0.2cm}\begin{lstlisting}[]
> OASIS.SetVoltageRange(2,2,3,3,4,4,5,5)
> OASIS.SetOversampling(2)
> OASIS.Sample(10,32000,0.1,0,TestFile)\end{lstlisting}\vspace{-0.2cm}
\end{OASISPrompt}

Those commands can either be issued using a serial interface, e.g., using the \textit{Serial Monitor} of the \textit{Arduino IDE} or by sending the commands as UDP packets. First, the voltage ranges are set with the corresponding ID (found in the command reference in \cref{sec:CommandRef}) to $\pm\SI{5}{\volt}$ for the first two channels, $\pm\SI{6.25}{\volt}$ for channel 3 \& 4, $\pm\SI{10}{\volt}$ for channel 5 \& 6 and $\pm\SI{12.5}{\volt}$ for the last two channels. Then, the oversampling factor is configured to x4, and the sampling is started for \SI{10}{\second} with a sampling rate of \SI{32}{\kilo\hertz} when the first channel surpasses the set trigger level of \SI{0.1}{\volt}. The results of the sample are saved in files named \texttt{TestFile}, and the corresponding metadata file contains the following:

\tikzset{external/export next=false}
\begin{fileContents}{TestFile.OASISmeta}
\lstset{style=fileContent}\vspace{-0.2cm}\begin{lstlisting}[]
t_sample,10.00;f_sample,32000.00;n_sample,320000;trigg_level,0.10;sync_mode,0;ADC_BITS,18;VoltageRangIDS,2,2,3,3,4,4,5,5;VoltageRanges,5.00,5.00,6.25,6.25,10.0,10.0,12.5,12.5;OversamplingFactor,x4;DeviceArchitecture,OASIS-UROS;DeviceName,OASIS-UROS Alpha;\end{lstlisting}\vspace{-0.2cm}
\end{fileContents}

Note that the \texttt{sync\_mode} field is a feature of the original \textit{OASIS} board and is not available at this time for \textit{OASIS-UROS}, but this option is kept for future releases. The sampled data can now be retrieved from the \texttt{.OASIShex} file on the SD card or using the \textit{OASIS-GUI}, detailed in \cref{sec:OperatingInstructions}.

The procedure of retrieving samples from the ADC and saving them to an SD card in real time is further detailed below. Further, the required logic to obtain the voltage values from the sample bits, and how the sampled data can be accessed for further usage is detailed. Lastly, the differences between normal sampling mode and triggered sampling, when a user-definable voltage level is exceeded, are explained.

\subsection{Retrieving individual samples from the ADC}

\begin{figure}[H]
	\centering
	\def\svgwidth{\textwidth}
	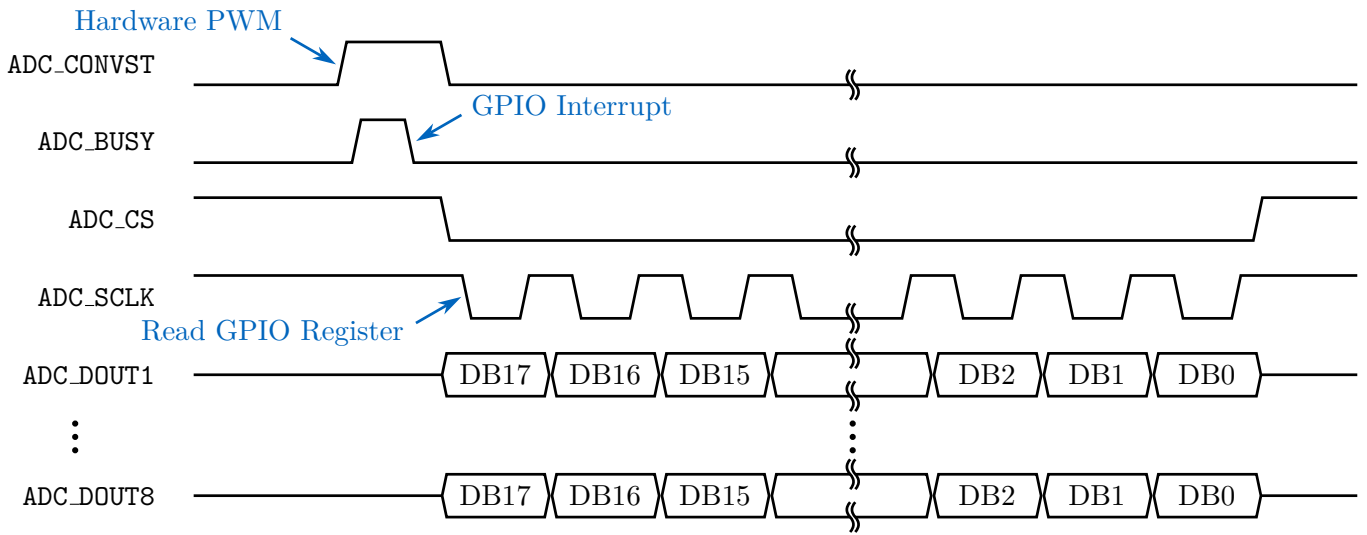\vspace{-0.4cm}
	\caption{Timing diagram of ADC sampling routine depicting logical values of pins over time, based on~\cite{ADCDatasheet}}
	\label{fig:adc_timing}
\end{figure}

The logic required to acquire one sample, consisting of one voltage value quantized using 18-bit for each of the eight channels, is depicted in \cref{fig:adc_timing}. This timing diagram, based on the ADC datasheet~\cite{ADCDatasheet}, shows the logical values, either zero or one, of the pins listed on the left, e.g., \texttt{ADC\_CONVST}.

\vspace{-0.4cm}

The sampling routine starts by requesting the ADC to sample the voltage currently applied to its input pins, which is achieved by setting \texttt{ADC\_CONVST} to \texttt{HIGH}. This transition is controlled by a hardware-generated PWM signal of the \textit{ESP32-S3} configured with the sampling frequency requested. After receiving the command to start the sampling process, the ADC will read the voltages on its input pins. During this time, the ADC signals that it is processing the sample by setting the \texttt{ADC\_BUSY} pin to \texttt{HIGH}.

One advantage of using a hardware-generated PWM signal instead of a timer, as used by the original \textit{OASIS} firmware, is a reduced jitter of the sample points. This is because timer interrupt service routines are scheduled by \textit{FreeRTOS}, and thus, the time when \texttt{ADC\_CONVST} is actually set to \texttt{HIGH} can vary. \textit{FreeRTOS} does not have to interact with the PWM signal. Further, with a timer interrupt the window to process samples would be between the falling \texttt{ADC\_BUSY} and the rising \texttt{ADC\_CONVST}, since the timer interrupt would interfere with reading the samples. Using PWM, the window increases to the time between two falling edges of \texttt{ADC\_BUSY} and effectively decouples the processing of the ADC and the \textit{ESP32-S3}, i.e., both can take almost the whole sample period to process. This allows for a higher oversampling factor to be set, increasing the signal-to-noise performance.

When the sample is ready, the ADC will transition the \texttt{ADC\_BUSY} pin to \texttt{LOW}. On this falling edge, a GPIO interrupt of the \textit{ESP32-S3} will be triggered that starts the process of reading the data lines into RAM using OctoSPI, an SPI implementation using eight data lines simultaneously. First, the chip select \texttt{ADC\_CS} is set to \texttt{LOW} to initiate the transaction. The falling edge of \texttt{ADC\_CS} leads to the ADC providing the most significant bit (MSB) of the sample on the data out lines. In \cref{fig:adc_timing}, the MSB is denoted by DB17, i.e., the 18th bit of the sample, where the MSB of channel 1 can be read from \texttt{ADC\_DOUT1}, for channel 2 on \texttt{ADC\_DOUT2} and so on.

\begin{figure}[H]
	\centering
	\def\svgwidth{\textwidth}
	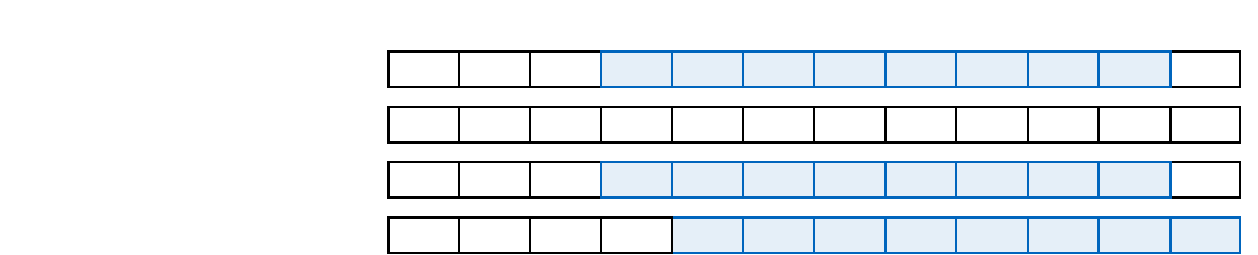\vspace{-0.4cm}
	\caption{Illustration of ADC sample extraction from the GPIO input register \texttt{GPIO\_IN\_REG} of the \textit{ESP32-S3}; marked in blue are the desired ADC sample bits}
	\label{fig:esp32_gpio_reg}
\end{figure}

The custom function \texttt{byte OASISDRIVER::readADCByte()} is used to read the values of the data out lines. At first, the clock \texttt{ADC\_SCLK} is set to low, which does not change the bits provided by the ADC. After this, the \SI{32}{\bit} of the first GPIO input register \texttt{GPIO\_IN\_REG} of the \textit{ESP32-S3} is read, see also the technical reference manual of the \textit{ESP32-S3}~\cite{ESP32S3TechRef}. The value of \texttt{ADC\_DOUT1} (GPIO1) is contained at the second least significant bit (LSB), \texttt{ADC\_DOUT8} (GPIO8) is the ninth LSB. Using a bit-wise \texttt{AND} with the appropriate bit mask \texttt{0x1FE}, followed by a bit-wise right shift of one, yields the desired eight bits, i.e., the first byte of the sample, see also \cref{fig:esp32_gpio_reg}.

\vspace{-0.4cm}

At the end of the custom function, the clock \texttt{ADC\_SCLK} is set back to \texttt{HIGH}, which signals the ADC to provide the next bits (DB16) on the data out lines.

This process is repeated until all eight bits of each channel, resulting in a sample size of 18 bytes, are retrieved. At the end, the chip select \texttt{ADC\_CS} is returned to \texttt{HIGH}, restoring the same logical levels as at the beginning of the sample procedure and the system waits for the next timer interrupt.

\subsection{Real-time sample acquisition and storage}

\begin{figure}[H]
	\centering
	\def\svgwidth{\textwidth}
	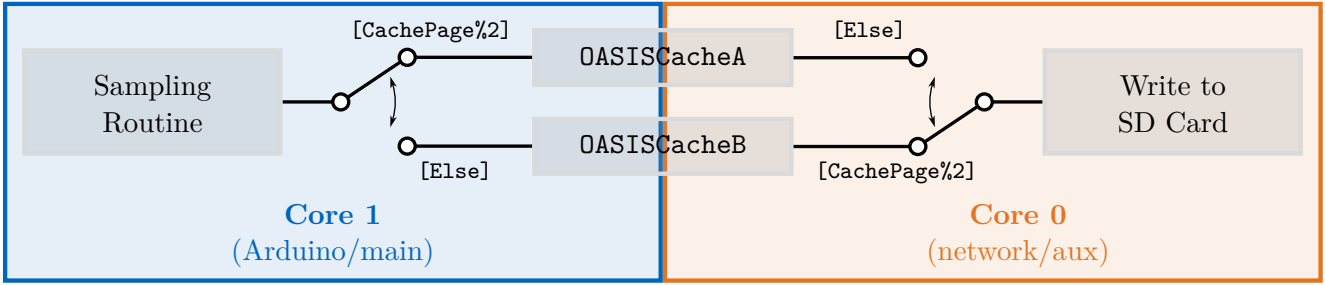\vspace{-0.4cm}
	\caption{Illustration of ADC sample transfer concept}
	\label{fig:cache_page_switch}
\end{figure}

Only crucial processing steps are performed to minimize the time spent processing the ADC samples during the data acquisition. Additionally, the second core of the \textit{ESP32-S3} is utilized. The main code, i.e., the Arduino sketch, by default runs on core 1, while core 0 handles all network-related code, e.g., the WiFi access point. This means that this core is shared with other tasks and is not suited for time-critical task, like retrieving samples from the ADC. However, it can be used for less critical tasks, like moving data to the SD card. This is the basis for the utilized sample data handling, see also \cref{fig:cache_page_switch}.

\vspace{-0.4cm}

The data from the ADC is stored in RAM in one of two arrays used for caching, \texttt{OASISCacheA} or \texttt{OASISCacheB}. Which cache is used is controlled by the variable \texttt{CachePage} that keeps track of how many caches have been filled and is initialized to 1. If the number is odd, i.e., the value of \texttt{CachePage} is not dividable by 2 and \texttt{CachePage\%2 = 1}, the data is written to \texttt{OASISCacheA}. If \texttt{CachePage} is odd, then \texttt{OASISCacheB} is used.

The position in the cache where the next sample should be written to is tracked using the variable \texttt{CacheIndex}, which is incremented by 1 after a new sample (18 bytes) has been written. When the last available position has been written to, the \texttt{CacheIndex} is reset to 0, and \texttt{CachePage} increased. Core 1 now saves new samples in the other cache, e.g., \texttt{OASISCacheB}, allowing core 0 to access the cache previously written to, e.g., \texttt{OASISCacheA}. At this point, core 0 is signaled to start writing data to the SD card by setting \texttt{PacketToWrite} to true.

The \texttt{CACHE\_SIZE} is chosen to be relatively large to reduce the overhead of writing to the SD card, e.g., from having to rewrite all data within one block. The only time constraint for the data write task is that it has to finish writing the cache to the SD card before core 1 fills the next cache page. Short interruptions by other tasks running on core 0 are therefore okay. When core 0 finishes writing the data, it acknowledges this by setting \texttt{PacketToWrite} to false. This also serves as a collision check. In case core 1 filled the next cache page and \texttt{PacketToWrite} is set to true, core 0 did not finish in time. Then, the sampling process is terminated, and an error is thrown, informing the user that the sample rate is probably too high.

The last and only partially filled cache page is handled at the end of the sample procedure. Using the variable \texttt{WriteLastPacket}, core 0 is informed to write the remaining data to the SD card, then the file handle is closed. Next, the samples written on the SD card are post-processed as described below.

\subsection{Post-processing of samples}

\begin{figure}[H]
	\centering
	\def\svgwidth{\textwidth}
	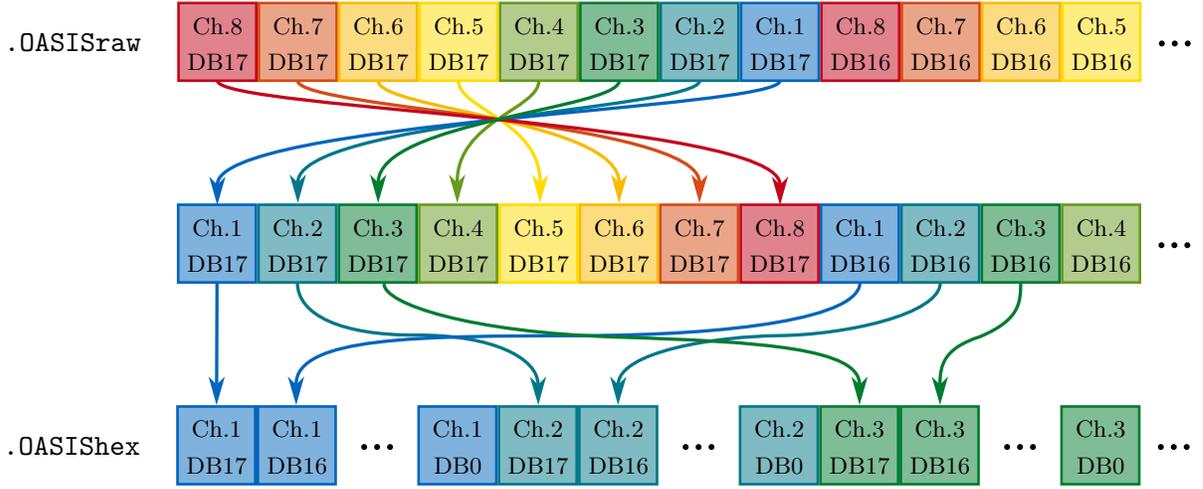\vspace{-0.4cm}
	\caption{Illustration of assembling the ADC data sorted by channel (\texttt{.OASIShex}) from sampled bits (\texttt{.OASISraw})}
	\label{fig:bit_sorting}
\end{figure}

After sampling, the samples are cached on the SD card in a \texttt{.OASISraw} file, written in the same order as they were acquired. For post-processing, one sample, i.e., the data of all eight channels at one moment in time, consisting of 18 bytes, is read into RAM. Then, the first bit in this chunk is the MSB (DB17) of channel 8, followed by the MSB of channel 7, etc., refer also to \cref{fig:bit_sorting}.

Two operations must to be performed to sort the data chunk into a usable format, i.e., where all bits belonging to one channel are consecutive and the channels are ordered from one to eight. First, the channels are separated into individual 32 bit variables, where the bits are moved to the correct positions, e.g., the MSB is bit 17. Then, to write the data back to the SD card, the channels are written into an 18 byte write cache, where the first byte contains DB17 through DB10 of channel 1, the second byte DB09 through DB02, and so on. The data formatted like this is written into the \texttt{.OASIShex} file. Even when the samples would be sent over the serial interface directly during the conversion process, this process is still necessary to maximize transfer speeds. Otherwise, the channel data would have to be zero-padded to 3 bytes, i.e., 24 bit, for the transfer. Since the sorting procedure is relatively simple, this is still done on the \textit{ESP32-S3} microcontroller. Using the Serial interface, the samples are transferred to a computer after sorting.

\begin{figure}[H]
	\centering
	\def\svgwidth{\textwidth}
	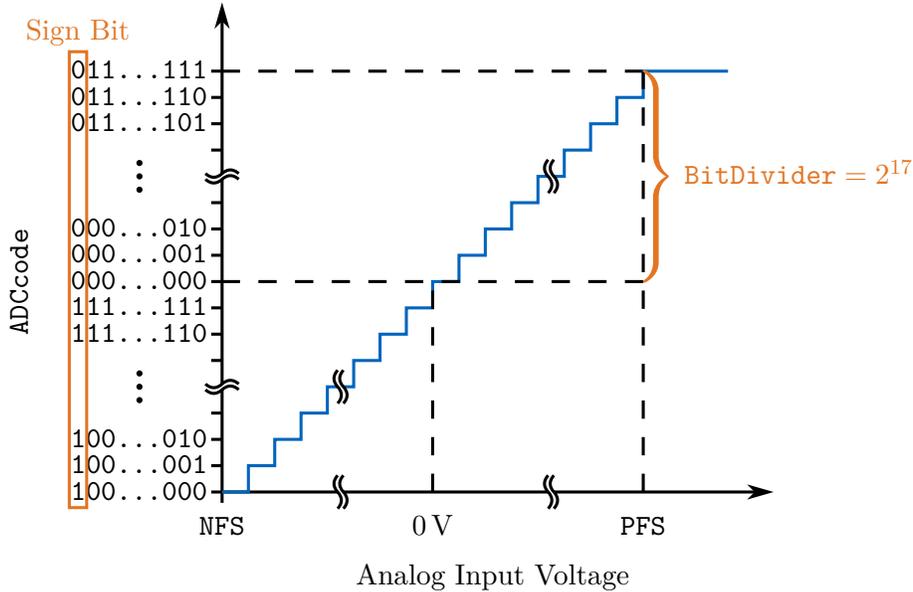\vspace{-0.4cm}
	\caption{Transfer function between the analog input voltage and \texttt{ADCCode}, based on~\cite{ADCDatasheet}}
	\label{fig:adc_bit_voltage}
\end{figure}

When the \textit{OASIS-GUI} is used, the samples are automatically converted into voltage values. For this, the bytes are sorted by channel again and buffered in a \textit{Numpy} array \texttt{OASISChannelData} with dimension of number of samples by 8. The algorithm to convert the \texttt{ADCCode}, i.e., the sorted bits of each sample and channel, is based on the ADC transfer function for bipolar input (positive and negative values) depicted in \cref{fig:adc_bit_voltage}.

\vspace{-0.4cm}

As can be derived from the figure, the first bit (MSB) is the sign bit, i.e., denotes whether the voltage is positive (MSB=0) or negative (MSB=1). The remaining bits indicate the magnitude of the voltage, where for MSB=0 all other bits equal to zero denotes a voltage of zero, and all bits equal to one denotes the positive full-scale (\texttt{PFS}) value, e.g., \SI{2.5}{\volt} for a voltage range of $\pm\SI{2.5}{\volt}$. For negative values (MSB=1), all other bits equal to zero indicate the negative full-scale (\texttt{NFS}), e.g., $-\SI{2.5}{\volt}$. Therefore, the measured voltage $V_\text{meas}$ is given by:
\begin{equation}
	V_\text{meas} = \begin{cases}
		\texttt{PFS}\cdot\dfrac{\texttt{ADCCode}}{\texttt{BitDivider}}\quad\text{for}\quad\text{MSB}=0\text{ (positive voltages)} \vspace{0.2cm}\\
		\abs{\texttt{NFS}}\cdot\dfrac{\texttt{ADCCode}-2\cdot\texttt{BitDivider}}{\texttt{BitDivider}}\quad\text{for}\quad\text{MSB}=1\text{ (negative voltages)} 
	\end{cases}
\end{equation}\vspace{-0.4cm}

To correctly convert the \texttt{ADCCode} to voltage, \texttt{PFS} and \texttt{NFS} must match the voltage range set for each channel. When the \textit{OASIS-GUI} is used to retrieve the samples, the required information was entered into the GUI and is known. In case \texttt{.OASIShex} files should be converted to voltages at a later point in time, the \texttt{.OASISmeta} file is required because it contains all parameters used for the data acquisition. Lastly, the time vector can be reconstructed using the sampling frequency and duration.

\subsection{Triggered sampling}

When a trigger is used to start the sampling process, there are a few differences that are described in this section. To capture the data before the sampling starts, i.e., the pre-trigger data, the system is continuously samples data after the command has been invoked. The sampled data is written into a third cache \texttt{OASISCachePre}, that is smaller than the caches used for sampling. The position within the cache is tracked using the variable \texttt{PreCacheIndex}. When the end of the cache is reached, \texttt{PreCacheIndex} is set back to zero, and the first data sampled is overwritten. This means that the system can remain in pre-trigger mode indefinitely, but also that the pre-trigger data is limited to as many samples as can fit into \texttt{OASISCachePre}. The size is defined in the firmware using \texttt{PRECACHE\_SIZE}, which is set to 1000 by default. Since the number of samples is fixed, the temporal length of the pre-trigger data will vary with the sample rate. Especially for low sample rates, the user must wait for the \texttt{OASISCachePre} to fill with data, otherwise, useless data is included in the samples.

To determine when the sampling process should start, defined by the user as a voltage level \texttt{trigg\_level} that channel 1 has to surpass, the samples of channel 1 are assembled and converted to voltage while sampling. When the set level is exceeded, the normal sampling procedure begins and ends with the transmission of the sorted bytes after the trigger. The pre-trigger data is sent after this data.

First, the pre-trigger data is written to the SD card in the file \texttt{$\ast$\_PRE.OASISraw}, where $\ast$ is the filename provided in the \texttt{OASIS.Sample()} command. Since \texttt{PreCacheIndex} holds the last index in the \texttt{OASISCachePre} that was written to, writing the data in the corrected order can be done in two steps. First, the data starting at \texttt{PreCacheIndex}+1 until the end is written, where \texttt{OASISCachePre}+1 is the oldest sample. Second, the data from the beginning of \texttt{OASISCachePre} until \texttt{PreCacheIndex} is written. The file \texttt{$\ast$\_PRE.OASISraw} is now in the correct temporal order, i.e., from oldest to newest sample, but still has to be post-processed as described above. Using the same procedures, the bits are sorted into \texttt{$\ast$\_PRE.OASIShex} and sent over the Serial interface.
% !TeX spellcheck = en_US
\section{Design files summary}

A summary of the provided design files is provided in the table below.
\vskip 0.1cm
\begin{table}[H]
	\tabulinesep=1ex
	\begin{tabu} to \linewidth {|X[1.25,1]|X|X|X|} 
		\hline
		\textbf{Design filename} & \textbf{File type} & \textbf{Open source license} & \textbf{Location of the file} \\\hline
		%Insert design files
		oasis-uros-hardware.zip & KiCad project & CC-BY 4.0 & \url{https://doi.org/10.5281/zenodo.13763227}~\cite{OASIS-UROS-V1.1-Zenodo} \\\hline
		jlcpcb\_manufacturing\_files.zip & JLCPCB manufacturing files & CC-BY 4.0 & \url{https://doi.org/10.5281/zenodo.13763227}~\cite{OASIS-UROS-V1.1-Zenodo} \\\hline
		oasis-firmware.zip & Arduino source code & MIT & \url{https://doi.org/10.5281/zenodo.13763227}~\cite{OASIS-UROS-V1.1-Zenodo} \\\hline
		oasis-gui.zip & Python package & MIT & \url{https://doi.org/10.5281/zenodo.13763227}~\cite{OASIS-UROS-V1.1-Zenodo} \\\hline
		bottom\_case.stl & STL file for 3D printing & CC-BY 4.0 & \url{https://doi.org/10.5281/zenodo.13763227}~\cite{OASIS-UROS-V1.1-Zenodo} \\\hline
	\end{tabu}
	\caption{Summary of design files provided in the repository}
\end{table}\vspace{-0.4cm}

\noindent\textbf{oasis-uros-hardware.zip}\hspace{0.2cm}This archive contains the full KiCad\footnote{\url{https://www.kicad.org/}} project of the \textit{OASIS-UROS} board as described here. Included are the KiCad schematics, the KiCad PCB, and all required symbols and footprints.

\noindent\textbf{jlcpcb\_manufacturing\_files.zip}\hspace{0.2cm}Included here are the production files used by JLCPCB to fabricate the partially assembled board. Using those files, the same board as used for validation can be ordered.

\noindent\textbf{oasis-firmware.zip}\hspace{0.2cm}In this archive, the source code for the board firmware can be found. \textit{OASIS-Firmware.ino} is the main code file; the other files are supplementary files. The code needs to be compiled using the Arduino IDE\footnote{\url{https://www.arduino.cc/en/software}}.

\noindent\textbf{oasis-gui.zip}\hspace{0.2cm}For archival purposes the source code for the \textit{OASIS-GUI} is included in the design files. However, it is recommended that the current version from the \textit{Python Package Index} is installed using \texttt{pip install OASIS-GUI}\footnote{\url{https://pypi.org/project/OASIS-GUI/}}. 

\noindent\textbf{bottom\_case.stl}\hspace{0.2cm}Lastly, for protection against mechanical damage,  shielding the bottom of the finished board is recommended. A 3D printing file for a simple case is included with this file.
% !TeX spellcheck = en_US
\section{Bill of materials summary}

Due to the size of some components, it is recommended to have the PCB at least partially assembled by the manufacturer. For the \textit{OASIS-UROS} board, we provide manufacturing files for JLCPCB that describe such a partially assembled board. This board can then be fully assembled as described in the build instructions. Following this approach, the table below lists all required components.
\vskip 0.2cm
\begin{table}[H]
	\tabulinesep=1ex
	\noindent{\footnotesize
	\begin{tabu} to \linewidth {|X[0.9,1]|X|X[0.4,1]|X[0.5,1]|X[0.6,1]|X[1.4,1]|X[0.75,1]|}
		\hline
		\textbf{Designator} & \textbf{Component} & \textbf{Number} & \textbf{Cost per unit - currency} & \textbf{Total cost - currency} & \textbf{Source of materials} & \textbf{Material type} \\\hline
		PCB & Printed \& partially assembled circuit board & 1 & $\approx$ \textdollar 35 & $\approx$ \textdollar 35 & \href{https://jlcpcb.com/}{JLCPCB} & Other \\\hline
		X1-X4 & Amphenol 031-6578 & 4 & \textdollar 14.25 & \textdollar 57.00 & \href{https://www.mouser.com/ProductDetail/523-31-6578}{523-31-6578 (Mouser)} & Metal \\\hline
		ADC0 & AD7606C-18BSTZ & 1 & \textdollar 51.35 & \textdollar 51.35 & \href{https://www.mouser.com/ProductDetail/584-AD7606C-18BSTZ}{584-AD7606C-18BSTZ (Mouser)} & Semiconductor \\\hline
		U50-U57 & LT3092EST & 8 & \textdollar 5.79 & \textdollar 46.32 & \href{https://www.mouser.com/ProductDetail/584-LT3092EST\%23TRPBF}{584-LT3092EST\#TRPBF (Mouser)} & Semiconductor \\\hline
		VR1-VR8 & 3296Y-1-203LF & 8 & \textdollar 2.42 & \textdollar 19.36 & \href{https://www.mouser.com/ProductDetail/652-3296Y-1-203LF}{652-3296Y-1-203LF (Mouser)} & Semiconductor \\\hline
		C40, C41 & MAL215099103E3 & 2 & \textdollar 2.33 & \textdollar 4.66 & \href{https://www.mouser.com/ProductDetail/594-MAL215099103E3}{594-MAL215099103E3 (Mouser)} & Semiconductor \\\hline
		L40 & SRR1260A-471K & 1 & \textdollar 0.78 & \textdollar 0.78 & \href{https://www.mouser.com/ProductDetail/652-SRR1260A-471K}{652-SRR1260A-471K (Mouser)} & Semiconductor \\\hline
		LS1 & PS1420P02CT & 1 & \textdollar 0.75 & \textdollar 0.75 & \href{https://www.mouser.com/ProductDetail/810-PS1420P02CT}{810-PS1420P02CT (Mouser)} & Semiconductor \\\hline
		JP1-JP8 & Jumper \& Pinheaders & 8 & \textdollar 0.23 & \textdollar 1.84 & \href{https://www.mouser.com/ProductDetail/571-28815452}{571-28815452 (Mouser)} & Other \\\hline
		
	\end{tabu}}
	\caption{Bill of materials summary using partially assembled PCB}
\end{table}\vspace{-0.4cm}
If a preassembly by the manufacturer is not desired, the table below lists the additional required components. Generic capacitors and resistors are not listed in the table below.

% JLCPCB Parts
\begin{table}[H]
	\tabulinesep=1ex
	\noindent{\footnotesize
	\begin{tabu} to \linewidth {|X[0.9,1]|X[1.2,1]|X[0.4,1]|X[0.5,1]|X[0.6,1]|X[1.2,1]|X[0.75,1]|}
		\hline
		\textbf{Designator} & \textbf{Component} & \textbf{Number} & \textbf{Cost per unit - currency} & \textbf{Total cost - currency} & \textbf{Source of materials} & \textbf{Material type} \\\hline
		U1 & ESP32-S3FN8 & 1 & \textdollar 3.2415 & \textdollar 3.2415 & \href{https://jlcpcb.com/partdetail/C2913196}{C2913196 (JLCPCB)} & Semiconductor \\\hline
		U40 & MC34063 & 1 & \textdollar 0.0881 & \textdollar 0.0881 & \href{https://jlcpcb.com/partdetail/C5349988}{C5349988 (JLCPCB)} & Semiconductor \\\hline
		U80 & USBLC6-2P6 & 1 & \textdollar 0.0974 & \textdollar 0.0974 & \href{https://jlcpcb.com/partdetail/C2827693}{C2827693 (JLCPCB)} & Semiconductor \\\hline
		U90 & TLV76733DRVR & 1 & \textdollar 0.2439 & \textdollar 0.2439 & \href{https://jlcpcb.com/partdetail/C2848334}{C2848334 (JLCPCB)} & Semiconductor \\\hline
		LED1-LED10 & WS2812C-2020-V1 & 10 & \textdollar 0.0626 & \textdollar 0.626 & \href{https://jlcpcb.com/partdetail/C2976072}{C2976072 (JLCPCB)} & Semiconductor \\\hline
		D40 & 1N5819HW-7-F & 1 & \textdollar 0.0279 & \textdollar 0.0279 & \href{https://jlcpcb.com/partdetail/C82544}{C82544 (JLCPCB)} & Semiconductor \\\hline
		F1 & JK-nSMD050-30 & 1 & \textdollar 0.03 & \textdollar 0.03 & \href{https://jlcpcb.com/partdetail/C720075}{C720075 (JLCPCB)} & Semiconductor \\\hline
		Y100 & SX2B40.000F1210F30 & 1 & \textdollar 0.0993 & \textdollar 0.0993 & \href{https://jlcpcb.com/partdetail/C2901733}{C2901733 (JLCPCB)} & Semiconductor \\\hline
		AE1 & RFANT3216120A5T & 1 & \textdollar 0.0654 & \textdollar 0.0654 & \href{https://jlcpcb.com/partdetail/C127629}{C127629 (JLCPCB)} & Semiconductor \\\hline
		T70 & DDTC114YCA-7-F & 1 & \textdollar 0.0273 & \textdollar 0.0273 & \href{https://jlcpcb.com/partdetail/C57530}{C57530 (JLCPCB)} & Semiconductor \\\hline
		SW1-SW3 & TS-1101-C-W & 3 & \textdollar 0.0348 & \textdollar 0.1044 & \href{https://jlcpcb.com/partdetail/C318938}{C318938 (JLCPCB)} & Semiconductor \\\hline
		J200 & U.FL-R-SMT-1(80) & 1 & \textdollar 0.0816 & \textdollar 0.0816 & \href{https://jlcpcb.com/partdetail/C88374}{C88374 (JLCPCB)} & Semiconductor \\\hline
		J4 & BM04B-SRSS-TB(LF)(SN) & 1 & \textdollar 0.1579 & \textdollar 0.1579 & \href{https://jlcpcb.com/partdetail/C160390}{C160390 (JLCPCB)} & Semiconductor \\\hline
		J5 & GT-TF003-H0185-02 & 1 & \textdollar 0.1272 & \textdollar 0.1272 & \href{https://jlcpcb.com/partdetail/C5155564}{C5155564 (JLCPCB)} & Semiconductor \\\hline
		J80 & GT-USB-7010ASV & 1 & \textdollar 0.0759 & \textdollar 0.0759 & \href{https://jlcpcb.com/partdetail/C2988369}{C2988369 (JLCPCB)} & Semiconductor \\\hline
	\end{tabu}}
	\caption{Summary of additionally required components when using bare PCB}
\end{table}\vspace{-0.4cm}
% !TeX spellcheck = en_US
\newpage
\section{Build instructions}

Using the provided design files, it is possible to have the acquisition board printed and fully assembled by the PCB manufacturer. However, there might be some minimum order quantities or issues with the availability of non-standard components. In this case, ordering an only partially assembled PCB, where only cheap and/or standard parts are placed, might be preferred. Using the manufacturing file provided for JLCPCB\footnote{\url{https://jlcpcb.com/}}, such a partially assembled PCB, as depicted in \cref{fig:JLCPCB_raw}, can be ordered.\vspace{-0.1cm}

\begin{figure}[H]
	\centering
	\def\svgwidth{\textwidth}
	%% Creator: Inkscape 1.3 (0e150ed6c4, 2023-07-21), www.inkscape.org
%% Accompanies image file 'jlcpcb_raw.pdf' (pdf, eps, ps)
\begingroup%
  \makeatletter%
  \providecommand\color[2][]{%
    \errmessage{(Inkscape) Color is used for the text in Inkscape, but the package 'color.sty' is not loaded}%
    \renewcommand\color[2][]{}%
  }%
  \providecommand\transparent[1]{%
    \errmessage{(Inkscape) Transparency is used (non-zero) for the text in Inkscape, but the package 'transparent.sty' is not loaded}%
    \renewcommand\transparent[1]{}%
  }%
  \providecommand\rotatebox[2]{#2}%
  \newcommand*\fsize{\dimexpr\f@size pt\relax}%
  \newcommand*\lineheight[1]{\fontsize{\fsize}{#1\fsize}\selectfont}%
  \ifx\svgwidth\undefined%
    \setlength{\unitlength}{5925.71673536bp}%
    \ifx\svgscale\undefined%
      \relax%
    \else%
      \setlength{\unitlength}{\unitlength * \real{\svgscale}}%
    \fi%
  \else%
    \setlength{\unitlength}{\svgwidth}%
  \fi%
  \global\let\svgwidth\undefined%
  \global\let\svgscale\undefined%
  \makeatother%
  \begin{picture}(1,0.53681272)%
    \lineheight{1}%
    \setlength\tabcolsep{0pt}%
    \put(0,0){\includegraphics[width=\unitlength,page=1]{jlcpcb_raw.pdf}}%
    \put(0.31650362,0.04985059){\color[rgb]{0.89019608,0.45882353,0.13333333}\makebox(0,0)[t]{\lineheight{1.25}\smash{\begin{tabular}[t]{c}$V_\text{BUS}$, \SI{5}{\volt}\end{tabular}}}}%
    \put(0.08934548,0.03257804){\color[rgb]{0.89019608,0.45882353,0.13333333}\makebox(0,0)[t]{\lineheight{1.25}\smash{\begin{tabular}[t]{c}$V_\text{logic}$, \SI{3.3}{\volt}\end{tabular}}}}%
    \put(0.23170921,0.49970583){\color[rgb]{0.89019608,0.45882353,0.13333333}\makebox(0,0)[t]{\lineheight{1.25}\smash{\begin{tabular}[t]{c}GND\end{tabular}}}}%
  \end{picture}%
\endgroup%
\vspace{-0.4cm}
	\caption{Partially assembled PCB using the provided JLCPCB manufacturing file (right) and recommended voltage supply rail testpoints \textbf{TP1}, \textbf{TP2} (left)}
	\label{fig:JLCPCB_raw}
\end{figure}\vspace{-0.4cm}

\noindent\textbf{Preliminary checks}\hspace{0.2cm} Before connecting the board to a PC, it is recommended to check the USB voltage rail for shorts to ground. For this, the resistance between testpoint \textbf{TP1} ($V_\text{BUS}$, \SI{5}{\volt}) to the chassis of the USB plug can be tested. Similarly, the output of the \SI{3.3}{\volt} supply can be tested for shorts using \textbf{TP2} ($V_\text{logic}$, \SI{3.3}{\volt}). If no shorts exist, the board can be connected to power via a USB-C cable, and \textbf{TP1} as well as \textbf{TP2} can be checked for the expected voltages. Note, that the USB specification allows for $V_\text{BUS}$ to be between 4.75 and \SI{5.25}{\volt}~\cite{USBImplementersForum2000}.

Then, the next step is to burn the bootloader for the \textit{ESP32-S3} microcontroller and flash the \textit{OASIS} firmware. Those steps are detailed in \cref{sec:SoftwareFlash}. If the board was ordered fully assembled, the steps after \cref{sec:SoftwareFlash} can be skipped until the current calibration, as described in \cref{sec:CurrentCalibration}. Otherwise, following the steps outlined below that detail how to assemble the PCB is recommended. \vspace{-0.15cm}

\subsection{Burning the bootloader and flashing the firmware}\label{sec:SoftwareFlash}\vspace{-0.15cm}

To burn the bootloader and flash the firmware, the Arduino IDE\footnote{\url{https://www.arduino.cc/en/software}} must be installed. Then, installing the board library 'esp32' by 'Espressif Systems' for the \textit{ESP32} family microcontroller through the \textit{Boards Manager} tab in the left column might be required. The software was developed and tested with version 3.0.3. Finally, one third-party library, 'Adafruit NeoPixel' by 'Adafruit', must be installed (tested with version 1.12.3).

After connecting the board to the PC with a USB-C cable, two settings must be selected before flashing the firmware. First, from the menu bar, 'Tools' $\rightarrow$ 'Board' $\rightarrow$ 'esp32' $\rightarrow$ 'ESP32S3 Dev Module' has to be selected. In the same menu, 'USB CDC On Boot' has to be set to 'Enabled', otherwise, there is no serial communication. Then, the firmware can be uploaded. If this is successful, the LED startup sequence, where the channel LEDs one through eight get lit in sequence, should play. Since there is no ADC, the communication with it fails and the system goes into an error state, signaled by a red blinking Status LED. When opening the 'Serial Monitor' (serial speed of 1,000,000 baud), the following output should be visible:

\tikzset{external/export next=false}
\begin{OASISPrompt}{OASIS Serial Interface - Startup Sequence}
	\lstset{style=OASISPrompt}\vspace{-0.2cm}\begin{lstlisting}[]
[OASIS] Open Acquisition System for IEPE Sensors starting up...

--------------------------------------------------------------------------
                          OASIS Device Information                          
--------------------------------------------------------------------------
[OASIS] Device Architecture is Unknown (ID 255)
[OASIS] Device Hardware Version is 255.255
[OASIS] Device Firmware Version is 255.255
[OASIS] ADC resolution: 255
[OASIS] TEDS module installed (1=yes, 0=no): 255
[OASIS] WSS module installed (1=yes, 0=no): 255
[OASIS] Device name: ������������������������
[OASIS] Current device configuration: MUTE: 1, WIFI_EN: 1
--------------------------------------------------------------------------

[OASIS] Resetting ADC...

[OASIS] Setting ADC data out format to 8 lines...
[OASIS] FATAL ERROR - Register write or readback failed!

[OASIS] Setting voltage range to +/- 2.5V for all channels...
[OASIS] Setting voltage ranges in ADC...
[OASIS] Successfully written 0x0 to address 0x3
[OASIS] Successfully written 0x0 to address 0x4
[OASIS] Successfully written 0x0 to address 0x5
[OASIS] Successfully written 0x0 to address 0x6
[OASIS] Voltage ranges set.

[OASIS] Setting oversampling factor x4 in ADC...
[OASIS] FATAL ERROR - Register write or readback failed!
[OASIS] Oversampling factor was NOT set.

[OASIS] Initializing SD Card...
E (5686) sdmmc_common: sdmmc_init_ocr: send_op_cond (1) returned 0x107
E (5687) vfs_fat_sdmmc: sdmmc_card_init failed (0x107).
[OASIS] FATAL ERROR - Failed to mount SD card!

[OASIS] FATAL ERROR - Critical self check failed!
\end{lstlisting}\vspace{-0.2cm}
\end{OASISPrompt}

\subsection{Soldering the ADC}\label{sec:SolderADC}

\begin{figure}[H]
	\centering
	\def\svgwidth{\textwidth}
	%% Creator: Inkscape 1.3 (0e150ed6c4, 2023-07-21), www.inkscape.org
%% Accompanies image file 'adc_assembly.pdf' (pdf, eps, ps)
\begingroup%
  \makeatletter%
  \providecommand\color[2][]{%
    \errmessage{(Inkscape) Color is used for the text in Inkscape, but the package 'color.sty' is not loaded}%
    \renewcommand\color[2][]{}%
  }%
  \providecommand\transparent[1]{%
    \errmessage{(Inkscape) Transparency is used (non-zero) for the text in Inkscape, but the package 'transparent.sty' is not loaded}%
    \renewcommand\transparent[1]{}%
  }%
  \providecommand\rotatebox[2]{#2}%
  \newcommand*\fsize{\dimexpr\f@size pt\relax}%
  \newcommand*\lineheight[1]{\fontsize{\fsize}{#1\fsize}\selectfont}%
  \ifx\svgwidth\undefined%
    \setlength{\unitlength}{5932.09261196bp}%
    \ifx\svgscale\undefined%
      \relax%
    \else%
      \setlength{\unitlength}{\unitlength * \real{\svgscale}}%
    \fi%
  \else%
    \setlength{\unitlength}{\svgwidth}%
  \fi%
  \global\let\svgwidth\undefined%
  \global\let\svgscale\undefined%
  \makeatother%
  \begin{picture}(1,0.48954057)%
    \lineheight{1}%
    \setlength\tabcolsep{0pt}%
    \put(0,0){\includegraphics[width=\unitlength,page=1]{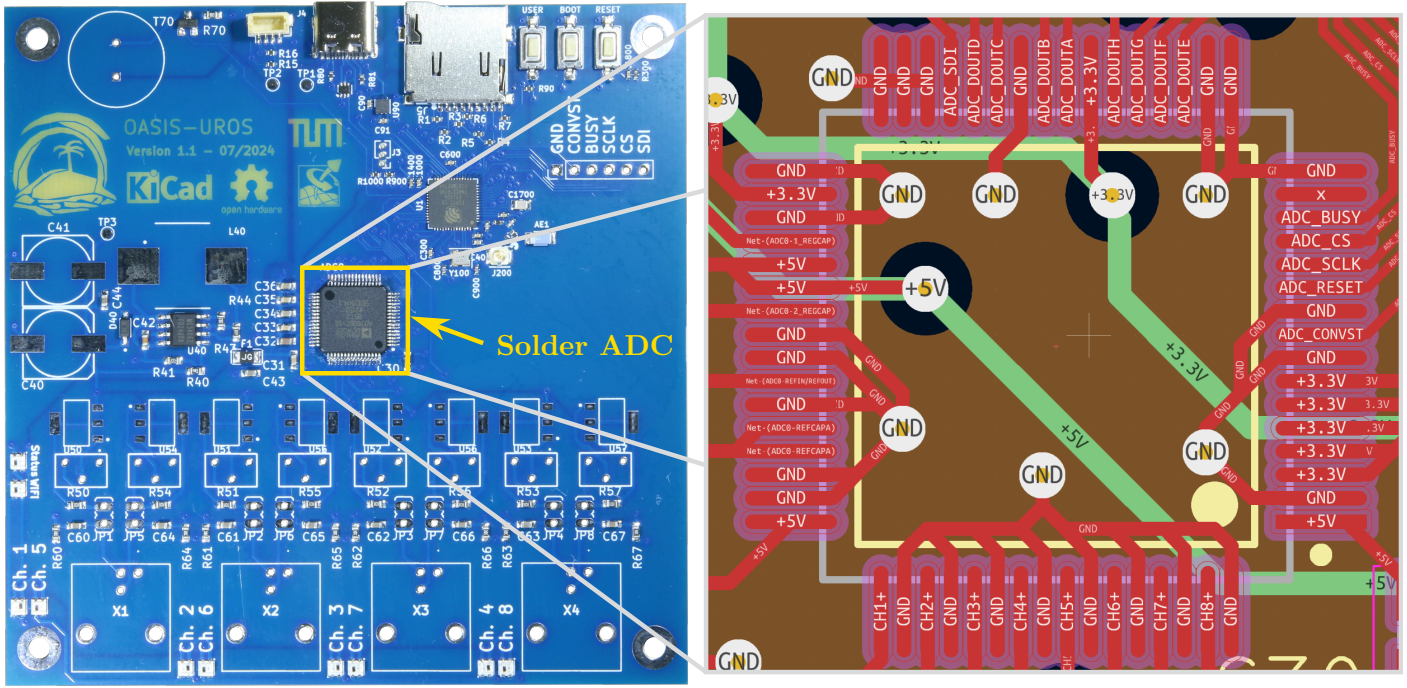}}%
    \put(0.35153642,0.23643668){\color[rgb]{1,0.8627451,0}\makebox(0,0)[lt]{\lineheight{1.25}\smash{\begin{tabular}[t]{l}\textbf{Solder ADC}\end{tabular}}}}%
  \end{picture}%
\endgroup%
\vspace{-0.4cm}
	\caption{Assembly step 1: Soldering the ADC \textbf{ADC0}, pinout shown on the right for quick reference}
	\label{fig:SolderADC}
\end{figure}

First, the ADC is soldered to the board since this is the most difficult part to solder due to the small footprint. The correct orientation of the ADC is shown on the left hand side of \cref{fig:SolderADC}. On the right hand side, the pinout of the ADC is shown for reference. Since some pins share the same signal, e.g., the three ground pins in the upper pin row on the left, some solder bridges between the pins might be okay. Using a drag soldering technique for the ADC and inspecting the pins for unwanted shorts under a microscope is recommended.

\vspace{-0.4cm}

\noindent\textbf{Check of ADC}\hspace{0.2cm}When connecting the board to the PC again, the following output should be displayed. As can be seen, for instance from \texttt{[OASIS] Successfully written 0x18 to address 0x2}, the communication with the ADC works as expected. If this is not the case, then there are some hardware defects.

\tikzset{external/export next=false}
\begin{OASISPrompt}{OASIS Serial Interface - Startup Sequence}
	\lstset{style=OASISPrompt}\vspace{-0.2cm}\begin{lstlisting}[]
[OASIS] Open Acquisition System for IEPE Sensors starting up...
        
        [...]
        
[OASIS] Resetting ADC...

[OASIS] Setting ADC data out format to 8 lines...
[OASIS] Successfully written 0x18 to address 0x2

[OASIS] Setting voltage range to +/- 2.5V for all channels...
[OASIS] Setting voltage ranges in ADC...
[OASIS] Successfully written 0x0 to address 0x3
[OASIS] Successfully written 0x0 to address 0x4
[OASIS] Successfully written 0x0 to address 0x5
[OASIS] Successfully written 0x0 to address 0x6
[OASIS] Voltage ranges set.

[OASIS] Setting oversampling factor x4 in ADC...
[OASIS] Successfully written 0x2 to address 0x8
[OASIS] Oversampling factor set.

[OASIS] Initializing SD Card...
E (5686) sdmmc_common: sdmmc_init_ocr: send_op_cond (1) returned 0x107
E (5687) vfs_fat_sdmmc: sdmmc_card_init failed (0x107).
[OASIS] FATAL ERROR - Failed to mount SD card!

[OASIS] FATAL ERROR - Critical self check failed!
	\end{lstlisting}\vspace{-0.2cm}
\end{OASISPrompt}

\subsection{Assembly of boost converter}\label{sec:SolderBoostConverter}

\begin{figure}[H]
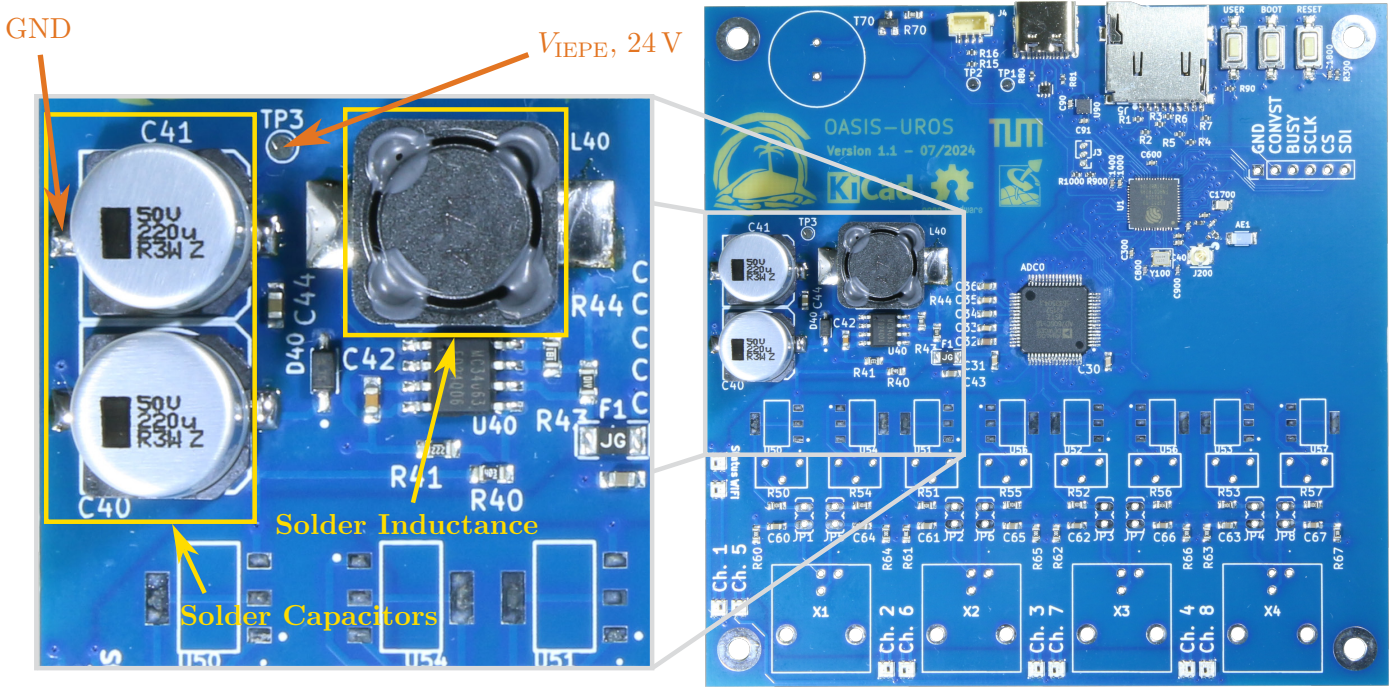

	\centering
	\def\svgwidth{\textwidth}
	%% Creator: Inkscape 1.3 (0e150ed6c4, 2023-07-21), www.inkscape.org
%% Accompanies image file 'bc_assembly.pdf' (pdf, eps, ps)
\begingroup%
  \makeatletter%
  \providecommand\color[2][]{%
    \errmessage{(Inkscape) Color is used for the text in Inkscape, but the package 'color.sty' is not loaded}%
    \renewcommand\color[2][]{}%
  }%
  \providecommand\transparent[1]{%
    \errmessage{(Inkscape) Transparency is used (non-zero) for the text in Inkscape, but the package 'transparent.sty' is not loaded}%
    \renewcommand\transparent[1]{}%
  }%
  \providecommand\rotatebox[2]{#2}%
  \newcommand*\fsize{\dimexpr\f@size pt\relax}%
  \newcommand*\lineheight[1]{\fontsize{\fsize}{#1\fsize}\selectfont}%
  \ifx\svgwidth\undefined%
    \setlength{\unitlength}{5926.71605869bp}%
    \ifx\svgscale\undefined%
      \relax%
    \else%
      \setlength{\unitlength}{\unitlength * \real{\svgscale}}%
    \fi%
  \else%
    \setlength{\unitlength}{\svgwidth}%
  \fi%
  \global\let\svgwidth\undefined%
  \global\let\svgscale\undefined%
  \makeatother%
  \begin{picture}(1,0.48981594)%
    \lineheight{1}%
    \setlength\tabcolsep{0pt}%
    \put(0.39162448,0.45347131){\color[rgb]{0.89019608,0.45882353,0.13333333}\makebox(0,0)[lt]{\lineheight{1.25}\smash{\begin{tabular}[t]{l}$V_\text{IEPE}$, \SI{24}{\volt}\end{tabular}}}}%
    \put(0,0){\includegraphics[width=\unitlength,page=1]{bc_assembly.pdf}}%
    \put(0.03487393,0.46370149){\color[rgb]{0.89019608,0.45882353,0.13333333}\makebox(0,0)[t]{\lineheight{1.25}\smash{\begin{tabular}[t]{c}GND\end{tabular}}}}%
    \put(0,0){\includegraphics[width=\unitlength,page=2]{bc_assembly.pdf}}%
    \put(0.13576789,0.04341899){\color[rgb]{1,0.8627451,0}\makebox(0,0)[lt]{\lineheight{1.25}\smash{\begin{tabular}[t]{l}\textbf{Solder Capacitors}\end{tabular}}}}%
    \put(0.29822297,0.1090801){\color[rgb]{1,0.8627451,0}\makebox(0,0)[t]{\lineheight{1.25}\smash{\begin{tabular}[t]{c}\textbf{Solder Inductance}\end{tabular}}}}%
  \end{picture}%
\endgroup%
\vspace{-0.4cm}
	\caption{Assembly step 2: Soldering the boost converter inductance \textbf{L40} and capacitors \textbf{C40}, \textbf{C41}, recommended testpoint \textbf{TP3} for $V_\text{IEPE}$ marked on the left}
	\label{fig:SolderBoostConverter}
\end{figure}

Next, the remaining components of the boost converter, the inductance \textbf{L40} as well as the two capacitors \textbf{C40} and \textbf{C41}, should be soldered as shown in \cref{fig:SolderBoostConverter}. Note that the capacitors have a fixed polarity and must be soldered in the correct orientation, respecting the outline drawn on the PCB. 

\vspace{-0.4cm}

\noindent\textbf{Check of the boost converter}\hspace{0.2cm}If everything is soldered correctly, the boost converter should be functional. This can be tested by connecting the board to power and measuring the voltage between testpoint \textbf{TP3} ($V_\text{IEPE}$, \SI{24}{\volt}) and GND, as indicated in \cref{fig:SolderBoostConverter}. Depending on the input voltage $V_\text{BUS}$ from the USB supply and the component tolerances, there might be some variations of the measured voltage. For the prototype, the open-circuit voltage of $V_\text{IEPE}$ was measured as \SI{23.948}{\volt} with an input voltage $V_\text{BUS}$ of \SI{4.9295}{\volt}\footnote{Measured using a Brymen BM786 Multimeter EEVBlog Edition in DC voltage mode; rated accuracy for utilized \SI{60}{\volt} range with 60,000 counts is $\pm\left(\SI{0.03}{\percent}\text{ + 2 digits}\right)$}.

\subsection{Assembly of constant current sources, buzzer, and BNC connectors}\label{sec:SolderCCSources}

\begin{figure}[H]
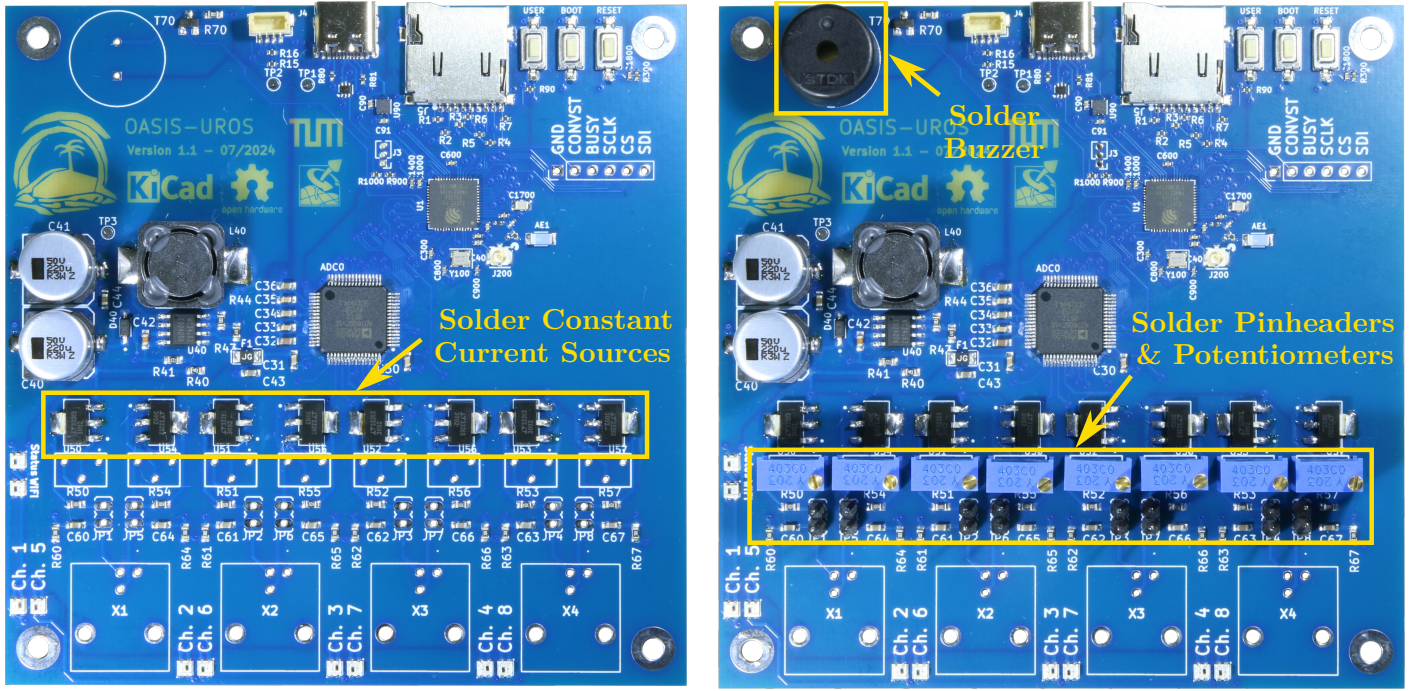

	\centering
	\def\svgwidth{\textwidth}
	%% Creator: Inkscape 1.3 (0e150ed6c4, 2023-07-21), www.inkscape.org
%% Accompanies image file 'cc_assembly.pdf' (pdf, eps, ps)
\begingroup%
  \makeatletter%
  \providecommand\color[2][]{%
    \errmessage{(Inkscape) Color is used for the text in Inkscape, but the package 'color.sty' is not loaded}%
    \renewcommand\color[2][]{}%
  }%
  \providecommand\transparent[1]{%
    \errmessage{(Inkscape) Transparency is used (non-zero) for the text in Inkscape, but the package 'transparent.sty' is not loaded}%
    \renewcommand\transparent[1]{}%
  }%
  \providecommand\rotatebox[2]{#2}%
  \newcommand*\fsize{\dimexpr\f@size pt\relax}%
  \newcommand*\lineheight[1]{\fontsize{\fsize}{#1\fsize}\selectfont}%
  \ifx\svgwidth\undefined%
    \setlength{\unitlength}{5926.71605869bp}%
    \ifx\svgscale\undefined%
      \relax%
    \else%
      \setlength{\unitlength}{\unitlength * \real{\svgscale}}%
    \fi%
  \else%
    \setlength{\unitlength}{\svgwidth}%
  \fi%
  \global\let\svgwidth\undefined%
  \global\let\svgscale\undefined%
  \makeatother%
  \begin{picture}(1,0.49167191)%
    \lineheight{1}%
    \setlength\tabcolsep{0pt}%
    \put(0,0){\includegraphics[width=\unitlength,page=1]{cc_assembly.pdf}}%
    \put(0.30615898,0.25713913){\color[rgb]{1,0.8627451,0}\makebox(0,0)[lt]{\lineheight{1.25}\smash{\begin{tabular}[t]{l}$\,$\textbf{Solder Constant}\vspace{-0.05cm}\\\textbf{Current Sources}\end{tabular}}}}%
    \put(0,0){\includegraphics[width=\unitlength,page=2]{cc_assembly.pdf}}%
    \put(0.80485017,0.25550478){\color[rgb]{1,0.8627451,0}\makebox(0,0)[lt]{\lineheight{1.25}\smash{\begin{tabular}[t]{l}\textbf{Solder Pinheaders}\vspace{-0.05cm}\\$\,$\textbf{\& Potentiometers}\end{tabular}}}}%
    \put(0,0){\includegraphics[width=\unitlength,page=3]{cc_assembly.pdf}}%
    \put(0.67080298,0.40322039){\color[rgb]{1,0.8627451,0}\makebox(0,0)[lt]{\lineheight{1.25}\smash{\begin{tabular}[t]{l}\textbf{$\,$Solder}\vspace{-0.05cm}\\\textbf{Buzzer}\end{tabular}}}}%
  \end{picture}%
\endgroup%
\vspace{-0.4cm}
	\caption{Assembly step 3: Soldering the constant current source ICs \textbf{U50}-\textbf{U57} (left) as well as the buzzer \textbf{LS1} (right), potentiometers \textbf{VR1}-\textbf{VR8} and pinheaders \textbf{JP1}-\textbf{JP8}}
	\label{fig:SolderCCSources}
\end{figure}

Now, the remaining components can be soldered. It is recommended to start with the last surface mount devices, the constant current sources \textbf{U50}-\textbf{U57}, as shown on the left of \cref{fig:SolderCCSources}. Then, the through-hole components can be assembled, starting with the buzzer \textbf{LS1} and then the potentiometers \textbf{VR1}-\textbf{VR8} as well as the pin headers \textbf{JP1}-\textbf{JP8}, see right of \cref{fig:SolderCCSources}.

\vspace{-0.4cm}

\begin{figure}[H]
	\centering
	\def\svgwidth{\textwidth}
	%% Creator: Inkscape 1.3 (0e150ed6c4, 2023-07-21), www.inkscape.org
%% Accompanies image file 'bnc_assembly.pdf' (pdf, eps, ps)
\begingroup%
  \makeatletter%
  \providecommand\color[2][]{%
    \errmessage{(Inkscape) Color is used for the text in Inkscape, but the package 'color.sty' is not loaded}%
    \renewcommand\color[2][]{}%
  }%
  \providecommand\transparent[1]{%
    \errmessage{(Inkscape) Transparency is used (non-zero) for the text in Inkscape, but the package 'transparent.sty' is not loaded}%
    \renewcommand\transparent[1]{}%
  }%
  \providecommand\rotatebox[2]{#2}%
  \newcommand*\fsize{\dimexpr\f@size pt\relax}%
  \newcommand*\lineheight[1]{\fontsize{\fsize}{#1\fsize}\selectfont}%
  \ifx\svgwidth\undefined%
    \setlength{\unitlength}{5926.71605869bp}%
    \ifx\svgscale\undefined%
      \relax%
    \else%
      \setlength{\unitlength}{\unitlength * \real{\svgscale}}%
    \fi%
  \else%
    \setlength{\unitlength}{\svgwidth}%
  \fi%
  \global\let\svgwidth\undefined%
  \global\let\svgscale\undefined%
  \makeatother%
  \begin{picture}(1,0.60454645)%
    \lineheight{1}%
    \setlength\tabcolsep{0pt}%
    \put(0,0){\includegraphics[width=\unitlength,page=1]{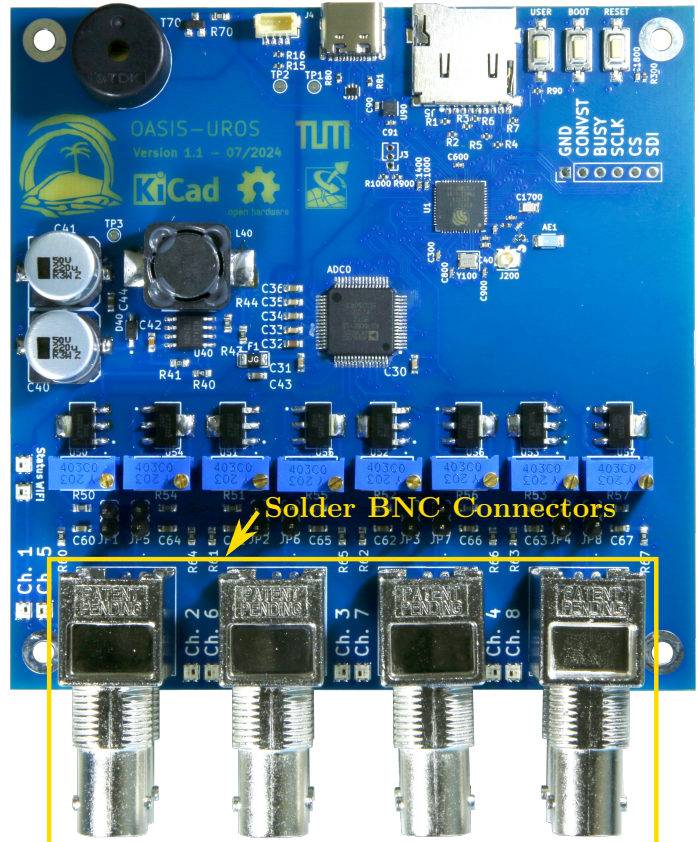}}%
    \put(0.43670029,0.23653){\color[rgb]{1,0.8627451,0}\makebox(0,0)[lt]{\lineheight{1.25}\smash{\begin{tabular}[t]{l}\textbf{Solder BNC Connectors}\end{tabular}}}}%
  \end{picture}%
\endgroup%
\vspace{-0.4cm}
	\caption{Assembly step 4: Soldering the BNC connectors \textbf{X1}-\textbf{X4}}
	\label{fig:SolderThruHole2}
\end{figure}

Lastly, the BNC connectors \textbf{X1}-\textbf{X4} should be soldered as depicted in \cref{fig:SolderThruHole2}. After this, the board is fully assembled. Before it can be used, the constant current sources must be calibrated as described next.

\vspace{-0.4cm}

\subsection{Current calibration}\label{sec:CurrentCalibration}

\begin{figure}[H]
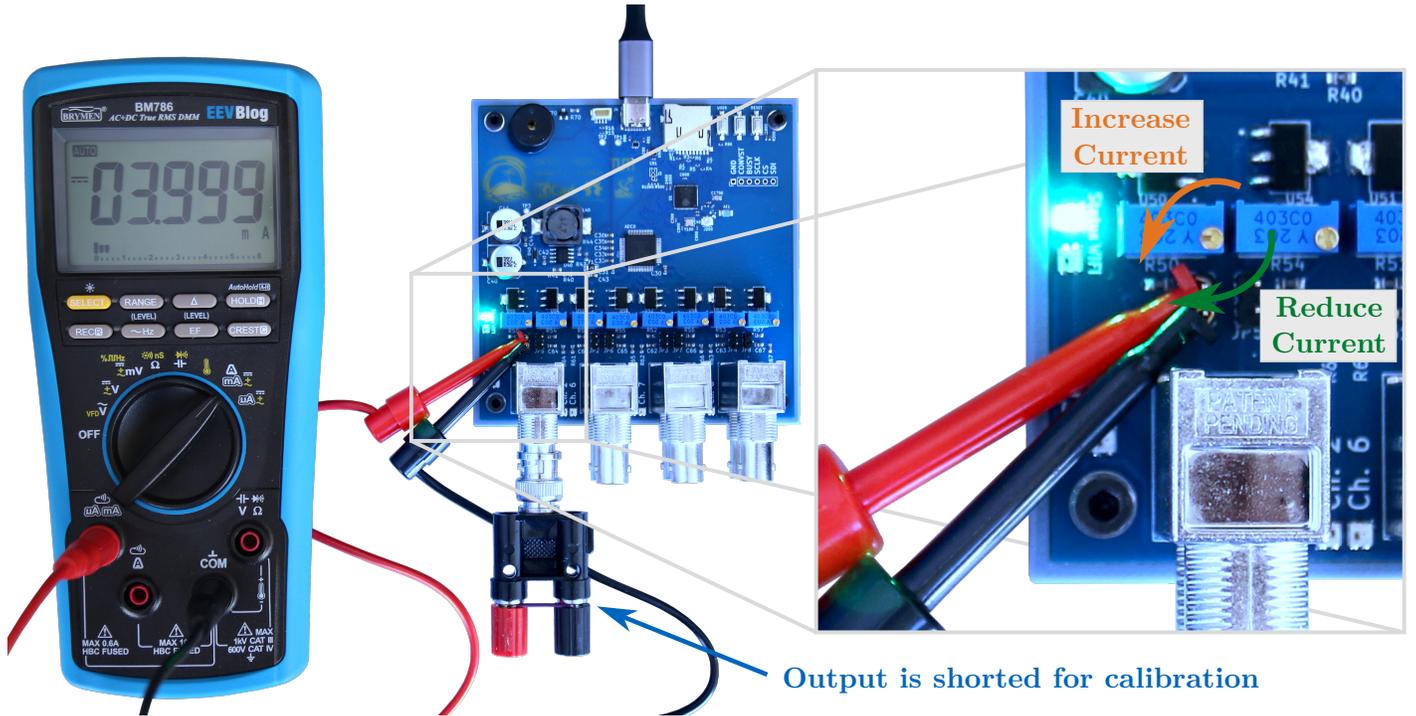

	\centering
	\def\svgwidth{\textwidth}
	%% Creator: Inkscape 1.3 (0e150ed6c4, 2023-07-21), www.inkscape.org
%% Accompanies image file 'current_calibration.pdf' (pdf, eps, ps)
\begingroup%
  \makeatletter%
  \providecommand\color[2][]{%
    \errmessage{(Inkscape) Color is used for the text in Inkscape, but the package 'color.sty' is not loaded}%
    \renewcommand\color[2][]{}%
  }%
  \providecommand\transparent[1]{%
    \errmessage{(Inkscape) Transparency is used (non-zero) for the text in Inkscape, but the package 'transparent.sty' is not loaded}%
    \renewcommand\transparent[1]{}%
  }%
  \providecommand\rotatebox[2]{#2}%
  \newcommand*\fsize{\dimexpr\f@size pt\relax}%
  \newcommand*\lineheight[1]{\fontsize{\fsize}{#1\fsize}\selectfont}%
  \ifx\svgwidth\undefined%
    \setlength{\unitlength}{6943.48878875bp}%
    \ifx\svgscale\undefined%
      \relax%
    \else%
      \setlength{\unitlength}{\unitlength * \real{\svgscale}}%
    \fi%
  \else%
    \setlength{\unitlength}{\svgwidth}%
  \fi%
  \global\let\svgwidth\undefined%
  \global\let\svgscale\undefined%
  \makeatother%
  \begin{picture}(1,0.50838951)%
    \lineheight{1}%
    \setlength\tabcolsep{0pt}%
    \put(0,0){\includegraphics[width=\unitlength,page=1]{current_calibration.pdf}}%
    \put(0.80314627,0.41994919){\color[rgb]{0.89019608,0.45882353,0.13333333}\makebox(0,0)[t]{\lineheight{1.25}\smash{\begin{tabular}[t]{c}\textbf{Increase}\\\textbf{Current}\end{tabular}}}}%
    \put(0.94497135,0.28489849){\color[rgb]{0,0.48627451,0.18823529}\makebox(0,0)[t]{\lineheight{1.25}\smash{\begin{tabular}[t]{c}\textbf{Reduce}\\\textbf{Current}\end{tabular}}}}%
    \put(0,0){\includegraphics[width=\unitlength,page=2]{current_calibration.pdf}}%
    \put(0.55385819,0.01965312){\color[rgb]{0,0.39607843,0.74117647}\makebox(0,0)[lt]{\lineheight{1.25}\smash{\begin{tabular}[t]{l}\textbf{Output is shorted for calibration}\end{tabular}}}}%
  \end{picture}%
\endgroup%
\vspace{-0.4cm}
	\caption{Recommended setup for calibrating the constant current sources of each channel}
	\label{fig:CurrentCalibration}
\end{figure}

To calibrate the constant current sources, an ammeter, e.g., found in a multimeter, and suitable test leads, e.g., pincer test leads as shown in \cref{fig:CurrentCalibration}, is required. Further, a load needs to be connected to the BNC headers, otherwise the output of the constant current source is open-circuit. For this, either an IEPE sensor can be connected, or, due to the current limitation through the IC, the constant current source output can simply be shorted to ground. One way to do this, is using a BNC to banana plug adapter and shorting the banana plugs using a wire, see \cref{fig:CurrentCalibration}. The latter approach is recommended to avoid potential damage to an expensive sensor, e.g., due to assembly errors. Then, the recommended approach is as follows:
\begin{enumerate}
	\item Connect the ammeter test leads to the pin header of the channel to be calibrated (polarity as in \cref{fig:CurrentCalibration})
	\item Short the BNC connector, for example, using a shorted BNC to banana plug adapter
	\item Adjust the current source using the corresponding potentiometer until the ammeter shows a current of \SI{4}{\milli\ampere}; the turning direction for increasing/decreasing the current is shown on the right of \cref{fig:CurrentCalibration}
	\item Remove the shorted BNC adapter and the test leads, then place a jumper on the pin header
\end{enumerate}

\vspace{-0.4cm}

Setting a current of \SI{4}{\milli\ampere} per channel is recommended for broad compatibility with IEPE devices~\cite{Lally2005}. Further, the boost converter was designed for this current draw ($8\cdot\SI{4}{\milli\ampere}=\SI{32}{\milli\ampere}$). As measurements of $V_\text{IEPE}$ for different current levels show, see \cref{tab:current_voltage}, the IEPE supply can provide \SI{32}{\milli\ampere} continuously without significant voltage drops, while at \SI{40}{\milli\ampere} (\SI{5}{\milli\ampere} per channel), the voltage drops.

\begin{table}[H]
	\centering
	\begin{tabular}{c|c|c|c|c|c|c}
		Current drawn & \SI{0}{\milli\ampere} & \SI{10}{\milli\ampere} & \SI{20}{\milli\ampere} & \SI{30}{\milli\ampere} & \SI{32}{\milli\ampere} & \SI{40}{\milli\ampere} \\\hline
		Measured $V_\text{IEPE}$ & \SI{23.948}{\volt} & \SI{23.928}{\volt} & \SI{23.919}{\volt} & \SI{23.918}{\volt} & \SI{23.918}{\volt} & \SI{22.604}{\volt}
	\end{tabular}
	\caption{Measured IEPE supply voltage $V_\text{IEPE}$ for various current draws after a settling time of approximately one hour with input voltage $V_\text{BUS}=\SI{4.9295}{\volt}$; measured using a Brymen BM786 Multimeter EEVBlog Edition in DC voltage mode; rated accuracy for utilized \SI{60}{\volt} range with 60,000 counts is $\pm\left(\SI{0.03}{\percent}\text{ + 2 digits}\right)$}
	\label{tab:current_voltage}
\end{table}

\subsection{Preparing the micro SD card and setting the device information}

First, a micro SD card should be formatted for use with the \textit{OASIS-UROS} board using a computer. The file system has to be \texttt{FAT32}, and, based on the utilized \texttt{CACHE\_SIZE} and some testing, a block size of \SI{32}{\kilo\byte} is recommended for the best performance. After formatting, insert the SD card into the board.

Now, the system should successfully start-up and complete all required self checks, indicated by the message \texttt{[OASIS] Finished booting.}. At this point, the device information should also be set by invoking \texttt{OASIS.SetDeviceInfo()}. This allows to set all device information stored in the EEPROM, see \cref{sec:CommandRef} for details, at once or individually. Important to correctly set is \texttt{ADC\_BIT}, required for correct conversion to voltages, and \texttt{OASIS\_VER}, which determines how the \textit{OASIS-GUI} interacts with the device.

After this, the board is fully assembled, configured, and ready for operation.
% !TeX spellcheck = en_US

\section{Operation instructions}\label{sec:OperatingInstructions}

In this section, the recommended operation using the \textit{OASIS-GUI} is described. The installation from the \textit{Python Package Index}\footnote{\url{https://pypi.org/project/OASIS-GUI/}} is described below, followed by the standard procedure to connect to the board, start a data acquisition, and retrieve a \texttt{.mat} file containing the measurement data.

\subsection{Installing the OASIS-GUI}

For this, a Python installation\footnote{\url{https://www.python.org/downloads/}} is required. To install the \textit{OASIS-GUI}, a command prompt/terminal with the Python installation registered in the \texttt{PATH} is needed. Then, the \textit{OASIS-GUI} can be installed using the \textit{Package Installer for Python (pip)}:

\tikzset{external/export next=false}
\begin{OASISPrompt}{Command Prompt/Terminal with Python in PATH}
	\lstset{style=OASISPrompt}\vspace{-0.2cm}\begin{lstlisting}[]
> pip install OASIS-GUI\end{lstlisting}\vspace{-0.2cm}
\end{OASISPrompt}

All required dependencies should be installed. The software was developed and tested under \textit{Windows}, however, due to the platform independence of Python and some compatibility testing, it is also possible to use the \textit{OASIS-GUI} under \textit{macOS} and \textit{Linux}.

\subsection{Connecting to the board}

To connect to an \textit{OASIS} board, it is sufficient to connect it to the PC using a USB-C cable and open the \textit{OASIS-GUI} from a command prompt/terminal with Python in the \texttt{PATH} using:

\tikzset{external/export next=false}
\begin{OASISPrompt}{Command Prompt/Terminal with Python in PATH}
	\lstset{style=OASISPrompt}\vspace{-0.2cm}\begin{lstlisting}[]
> oasis-gui\end{lstlisting}\vspace{-0.2cm}
\end{OASISPrompt}

\begin{figure}[H]
	\centering
	\def\svgwidth{\textwidth}
	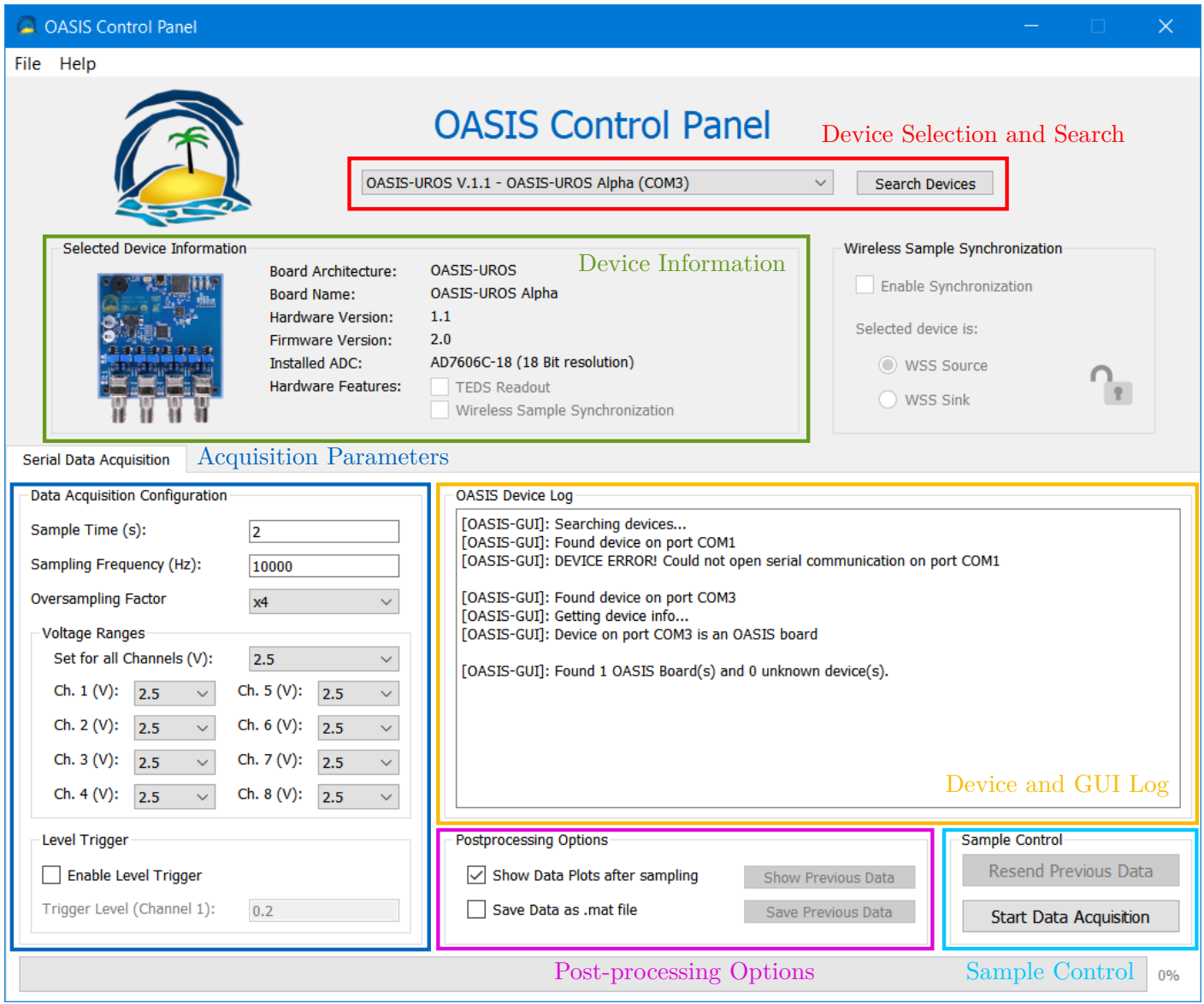\vspace{-0.4cm}
	\caption{\textit{OASIS-GUI} after opening with an \textit{OASIS-UROS} board connected}
	\label{fig:oasis_gui_device_search}
\end{figure}

This opens the user interface, see \cref{fig:oasis_gui_device_search}, and searches for connected devices. By attempting to open communication with every connected Serial device and sending a request for device information (\texttt{OASIS.RawInfo()}), the connected boards and their COM port can be determined. All \textit{OASIS} boards are placed into the drop-down menu at the top in the \textit{Device Selection and Search} segment. The segments are explained in the following.

\newpage

\vspace{-0.4cm}

\noindent\textbf{Device Selection and Search}\hspace{0.2cm}Besides selecting the active device, i.e., the one used for data acquisition, this section allows to re-scan the connected devices using the \textit{Search Devices} button.

\noindent\textbf{Device Information}\hspace{0.2cm}Here, the information stored in the device EEPROM (see \cref{sec:CommandRef}) is listed. 

\noindent\textbf{Acquisition Parameters}\hspace{0.2cm}These parameters are used for the next sample and are populated with default values. The \textit{OASIS-GUI} sets the voltage ranges and the oversampling factor before the sample begins. If the sample should start with a trigger, this can also be enabled here, at the bottom of the window.

\noindent\textbf{Device and GUI Log}\hspace{0.2cm}Logs from the connected \textit{OASIS} devices and the \textit{OASIS-GUI} are displayed here. The latter contains, for example, information about found Serial devices and if the communication was successful.

\noindent\textbf{Post-processing Options}\hspace{0.2cm}Here, the behavior after a completed sample can be configured. It can be selected whether the acquired data should be plotted or saved after the sample. Using the dedicated buttons next to the options, both options can be manually invoked later.

\noindent\textbf{Sample Control}\hspace{0.2cm}The data acquisition can be started using the \textit{Start Data Acquisition} button. In case of transmission errors, the previous sample can be resend using \textit{Resend Previous Data}.

\newpage

\subsection{Sampling data}

Before starting the sampling process, the data acquisition parameters should be set. In the interest of flexibility, and to allow some experimentation, the sample parameters are as unlimited as possible. This means that there are some combinations of sample rate and oversampling factor that will not work. Generally, either a high sample rate (up to \SI{36}{\kilo\hertz}) or a high oversampling factor is possible. In the following, the individual parameters and their limitations are quickly discussed.

\noindent\textbf{Sample Time (s)}\hspace{0.2cm}Defines for how long samples are acquired in seconds. It is also possible to enter decimal values, e.g., 2.56. There is no limitation for the sample time, except the storage space of the SD card.

\noindent\textbf{Sampling Frequency (Hz)}\hspace{0.2cm}Determines how many samples are taken per second and must be a whole number. Due to limitations of the \textit{ESP32-S3} PWM hardware, not all values are possible, especially low sampling frequencies like \SI{10}{\hertz}. When this is the case, there will be an error message in the device log, and the sample process is aborted. As an upper limit, a sampling frequency of \SI{36}{\kilo\hertz} (with x16 oversampling) was found. If the sampling frequency is too high, i.e., the processing cannot be completed in time, the \textit{OASIS} firmware will detect this and cancel the data acquisition with an error message. This message also denotes whether retrieving the samples from the ADC or writing the cache pages to the SD card took too long.

\noindent\textbf{Oversampling Factor}\hspace{0.2cm}To reduce noise, the ADC can take multiple samples in quick succession that are averaged and this average is provided as the sample. How many samples are averaged is determined by the oversampling factor, e.g., a oversampling factor of x4 means that 4 samples are averaged~\cite{ADCDatasheet}. Only certain values, listed in \cref{tab:OSOASIS}, can be chosen from the drop-down menu.

\begin{table}[h!]
	\centering
	\begin{tabular}{c|c|c|c|c|c|c|c|c|c} 
		Oversampling factor & x1 (Off) & x2 & x4 & x8 & x16 & x32 & x64 & x128 & x256 \\\hline
		Theoretical maximum throughput (kSPS) & 1000 & 500 & 250 & 125 & 62.5 & 31.25 & 15.6 & 7.8 & 3.9
	\end{tabular}
	\caption{Oversampling factors and maximum theoretical throughput in kilo samples per second (kSPS)~\cite{ADCDatasheet}}
	\label{tab:OSOASIS}
\end{table}

In \cref{tab:OSOASIS}, the theoretical maximum throughput is given with respect to the ADC's capabilities. Due to other processing times, the achievable throughput, and with that, the maximum oversampling factor, is lower. Setting an oversampling factor that is too high can lead to unexpected results.

\noindent\textbf{Voltage Ranges}\hspace{0.2cm}Here, the bipolar voltage range (positive and negative values) can be set for each channel individually using the drop-down menus. The \textit{Set for all Channels (V)} option allows to set all eight channels to a specific range at once, channels can then still be changed individually. \Cref{tab:VROASIS} lists all available voltage ranges and the corresponding voltage resolution achievable.

\begin{table}[h!]
	\centering
	\begin{tabular}{c|c|c|c|c|c} 
		Voltage range  & $\pm\SI{2.5}{\volt}$ & $\pm\SI{5}{\volt}$ & $\pm\SI{6.25}{\volt}$ & $\pm\SI{10}{\volt}$ & $\pm\SI{12.5}{\volt}$ \\\hline
		Voltage resolution & \SI{19}{\micro\volt} & \SI{38.1}{\micro\volt} & \SI{47.7}{\micro\volt} & \SI{76.3}{\micro\volt} & \SI{95.36}{\micro\volt}
	\end{tabular}
	\caption{Available voltage ranges and corresponding resolution~\cite{ADCDatasheet}}
	\label{tab:VROASIS}
\end{table}

\noindent\textbf{Level Trigger}\hspace{0.2cm}This section allows to configure a voltage trigger on the first channel. When \textit{Enable Level Trigger} is checked, the trigger level can be entered as a decimal voltage. The pre-trigger length cannot be set by the user and is predefined in the \textit{OASIS} firmware through \texttt{PRECACHE\_SIZE}, which is set to 1000 samples.

When all parameters are chosen as desired, the sample process can be started using the \textit{Start Data Acquisition} button. During the data acquisition, the current status can be followed through the \textit{OASIS Device Log}. The status bar at the bottom denotes the progress of transmitting the samples to the computer. For \textit{OASIS-UROS}, this bar does not progress until after the sample has been completed and post-processed.

\begin{figure}[H]
	\centering
	\def\svgwidth{\textwidth}
	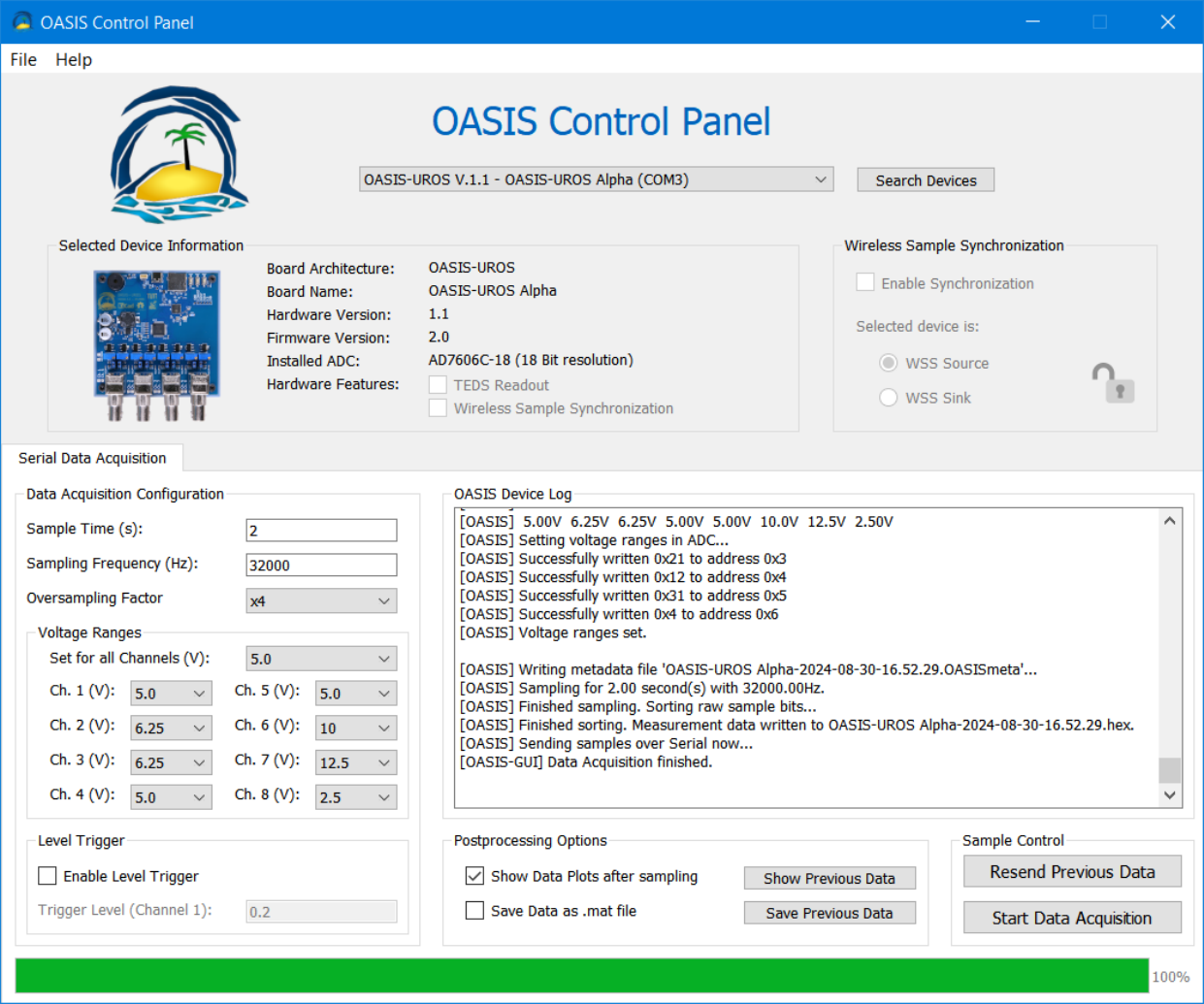\vspace{-0.4cm}
	\caption{\textit{OASIS-GUI} after data acquisition completed successfully}
	\label{fig:oasis_gui_sampled}
\end{figure}

After successful data acquisition, the \textit{OASIS-GUI} looks as displayed in \cref{fig:oasis_gui_sampled}. Note that the device log also displays the name of the sample, here \texttt{OASIS-UROS Alpha-2024-08-30-16.52.29}, containing the device name as well as the date and time of the sample (based on the computer's clock).

\vspace{-0.4cm}

\begin{figure}[H]
	\centering
	\includegraphics[width=0.97\textwidth]{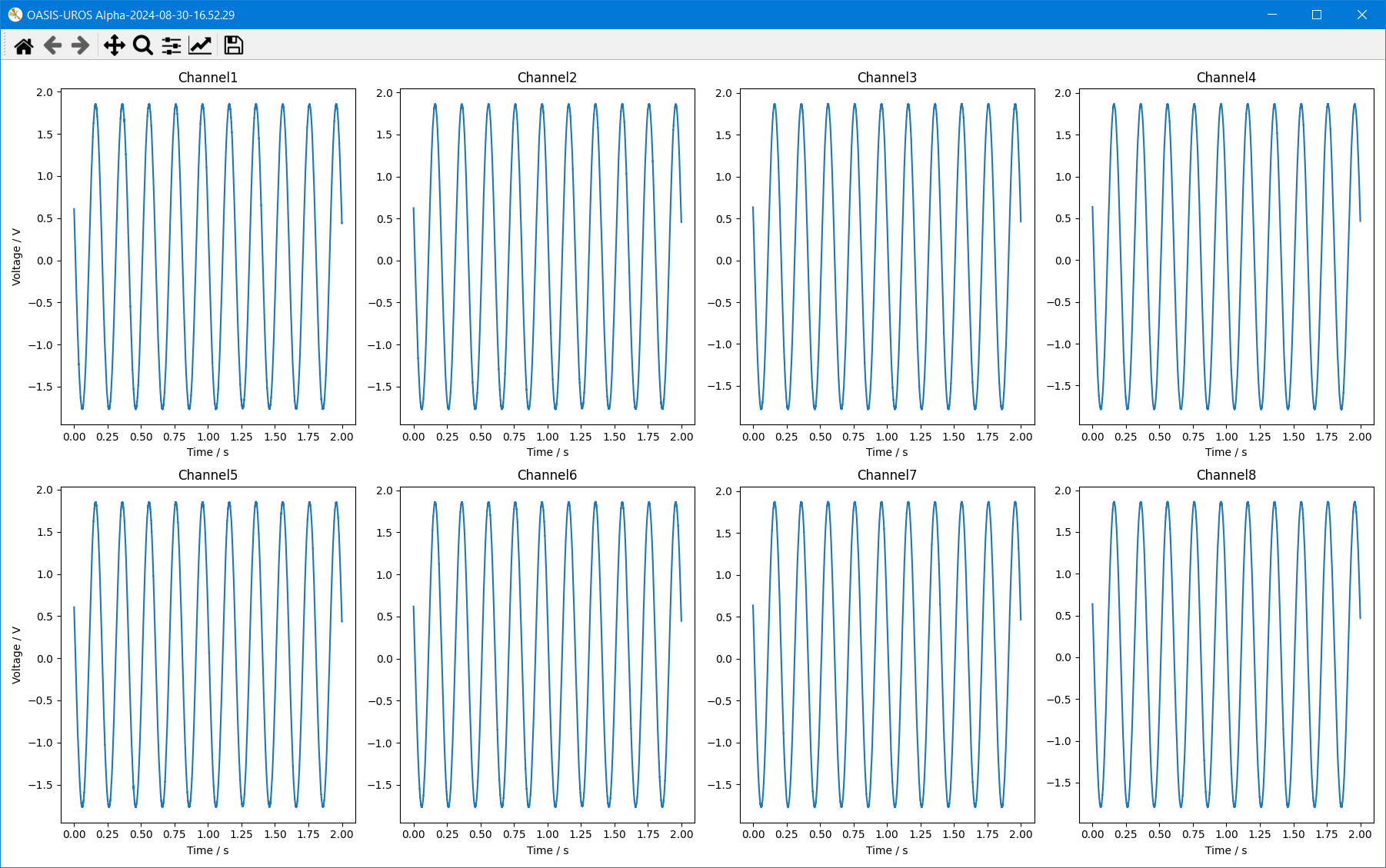}
	\caption{Plot displayed by \textit{OASIS-GUI} after the data acquisition is completed}
	\label{fig:oasis_gui_normal_plot}
\end{figure}

\begin{figure}[H]
	\centering
	\includegraphics[width=0.97\textwidth]{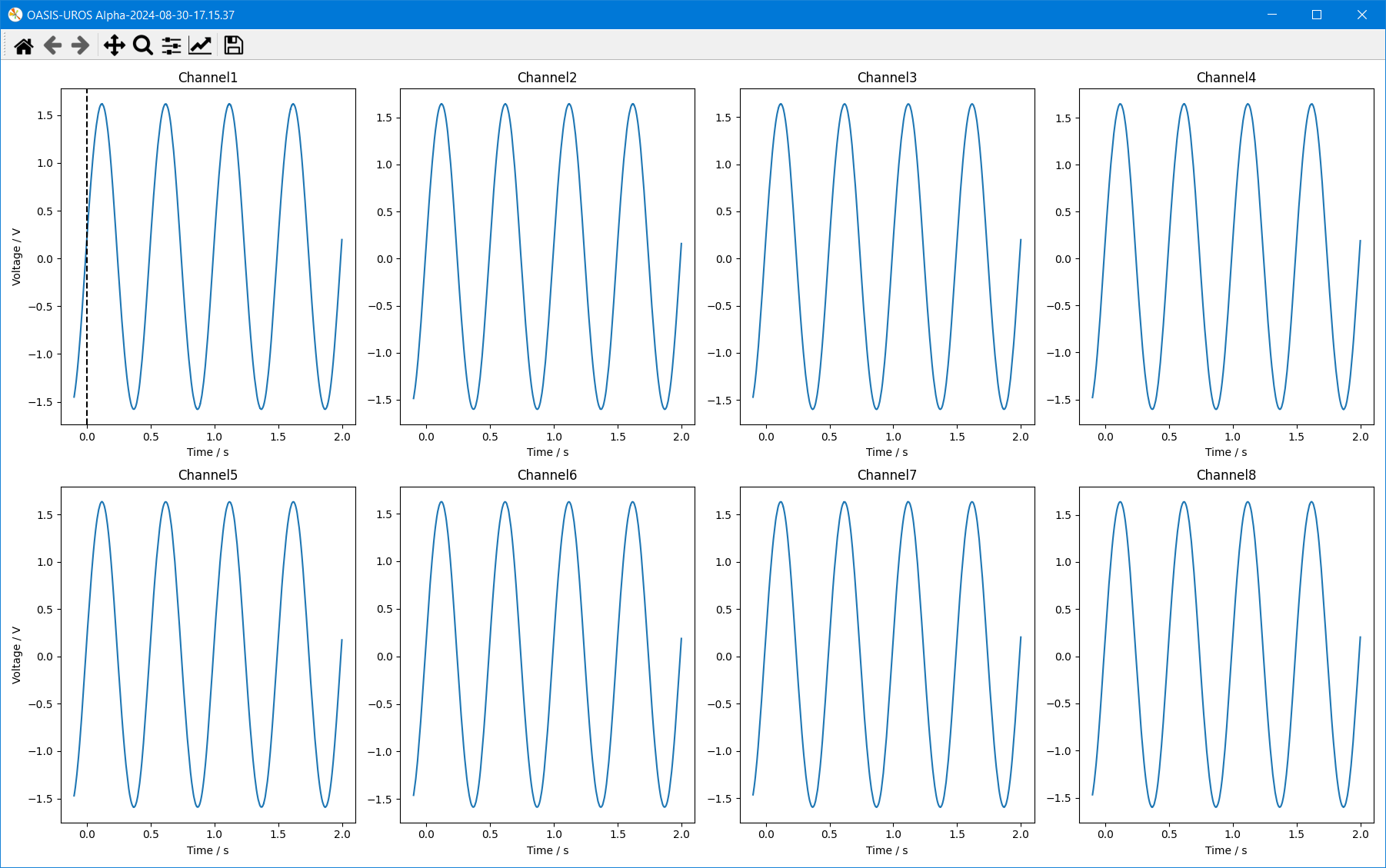}
	\caption{Plot displayed by \textit{OASIS-GUI} after the triggered data acquisition is completed}
	\label{fig:oasis_gui_triggered_plot}
\end{figure}

If not already selected as a post-processing option, the user can now manually invoke a plot of the sample with \textit{Show Previous Data}. This will create a \textit{matplotlib} plot of all eight channels over time, as depicted in \cref{fig:oasis_gui_normal_plot}. If the sample was started with a trigger, the plot will look like shown in \cref{fig:oasis_gui_triggered_plot}. The difference is that the time axis does not start at zero, but at a negative value, denoting the pre-trigger data. Located at $t=0$ is the first sample that exceeded the set trigger level, which is visualized with a vertical dotted line as depicted in \cref{fig:oasis_gui_triggered_plot}.

Sometimes, there can be errors in the transmission from the \textit{OASIS} board to the computer. This manifests in obvious jumps in all channels at the same time. However, this can be remedied by requesting the sampled data again using the \textit{Resend Previous Data} button.

Additionally, the user can save this data in a \texttt{.mat} file with the name of the sample using the \textit{Save Previous Data} button. The \texttt{.mat} file will contain two variables:
\begin{itemize}
	\item \texttt{OASISChannel} (8 $\times$ number of samples): Converted voltage sample points for each channel
	\item \texttt{OASISTime} (1 $\times$ number of samples): Corresponding time axis
\end{itemize}

\newpage

\vspace{-0.4cm}
\vspace{-0.4cm}

\newpage
% !TeX spellcheck = en_US

\section{Validation and characterization}\vspace{-0.05cm}

The performance and overall usability of the developed system are tested in the context of an experimental modal analysis. For this, a test case is set up and measurements are performed with the \textit{OASIS-UROS} system as well as a commercial system, the \textit{Siemens LMS Scadas} system. From the measured time data, frequency response functions (FRFs) and modal parameters (eigenfrequencies, damping ratios, and mode shapes) are estimated and compared using open source Python packages (\textit{pyFRF}~\cite{pyFRF} and \textit{pyFBS}~\cite{pyFBS}) for the \textit{OASIS} measurements and the proprietary software \textit{Simcenter Testlab 2406} for the \textit{LMS} measurements.\vspace{-0.05cm}

\subsection{Experimental test setup}\vspace{-0.05cm}

\begin{figure}[H]
	\centering
	\def\svgwidth{\textwidth}
	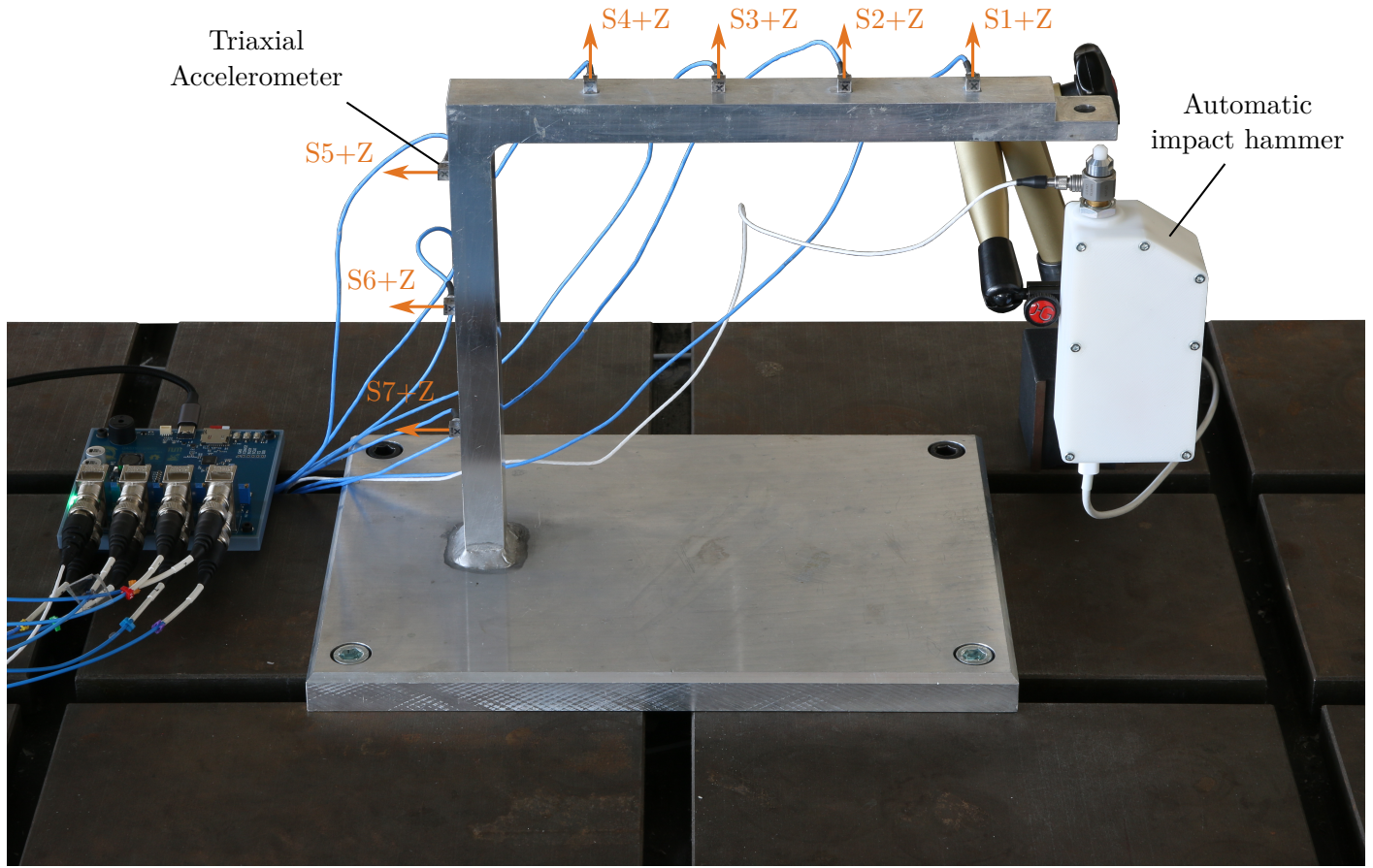\vspace{-0.4cm}
	\caption{Experimental setup of stiff aluminum beam structure fixed to a vibration-isolated table. The system is excited using an automatic impact hammer; acceleration responses perpendicular to the surface are measured using triaxial accelerometers, once with a commercial system and once using \textit{OASIS-UROS}}
	\label{fig:beam_structure}
\end{figure}

As a test case, a stiff aluminum beam structure is used, as depicted in \cref{fig:beam_structure}. Using screws, the structure is fixed to a vibration-isolated table. The system is excited with an automatic impact hammer provided by \textit{Maierhofer-Technology}\footnote{\url{https://www.maierhofer-technology.de}}. Seven acceleration responses are measured using triaxial acceleration sensors (\textit{PCB Piezotronics Model 356A03}) glued to the structure. Only the vibrations perpendicular to the surface ($z$-direction in the sensor coordinate system, as visualized in \cref{fig:beam_structure}) are measured.

Five impacts are performed using the same experimental setup, once with the sensors connected to the \textit{LMS} system and once connected to the \textit{OASIS} system. While this way, the signals will not be identical between the systems, it allows to validate the full electronic circuitry of \textit{OASIS-UROS}, i.e., including the IEPE supply.

\vspace{-0.4cm}

\newpage

For both systems, the measurements were started using a trigger on the impact force and a set measurement duration of \SI{2.56}{\second}. While for the \textit{LMS} system, the pre-trigger is included in this time, for \textit{OASIS}, the pre-trigger is acquired additionally, resulting in \SI{40}{\milli\second} of additional time data. A sample rate of \SI{25.6}{\kilo\hertz} was used for \textit{OASIS} and \SI{51.2}{\kilo\hertz} for \textit{LMS}. The following analysis is limited to a maximum frequency of \SI{12}{\kilo\hertz}.

The raw time data of both systems, as well as the FRFs calculated using \textit{Simcenter Testlab}, are available for download at~\cite{OASISUROSData}. Additionally, the \textit{Python} script used to calculate FRFs and fit modal parameters is provided.

\subsection{Frequency response function measurements}

\begin{figure}[H]
	\centering
	\input{figures/validation/excitation_FFT.tikz}\vspace{-0.25cm}
	\caption{Excitation signal of the automatic impulse hammer for each impact over time (left) and frequency (right). Impact belongs to data set of: \includegraphics{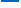}$\,$\textit{Siemens LMS Scadas} System, \includegraphics{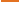}$\,$\textit{OASIS-UROS} System}
	\label{fig:excitation_FFT}
\end{figure}

Since individual impacts are performed for each system, and due to some variations while changing the acquisition system, there are also differences in the excitation characteristics. To compare the impacts, \cref{fig:excitation_FFT} shows  all performed impacts' time signal and excitation bandwidth.

\vspace{-0.4cm}

As can be seen, there is a drop in force magnitude between the individual runs. Compared to the impacts measured by the \textit{LMS} system, the impacts performed for the \textit{OASIS} measurements are overall lower in magnitude. Further, the excitation bandwidth is slightly lower for \textit{OASIS}, which also manifests in a significant dip around \SI{6}{\kilo\hertz} for some of the measurements.

Before the FRF calculations, a force window (equal to one near the impact, zero elsewhere) is applied to the excitation time series of both systems. No response window was required for the \textit{LMS} system, while the \textit{OASIS} system profited from applying an exponential window, resulting in cleaner FRFs. Because the energy loss due to the window is not corrected, the magnitudes of the \textit{OASIS} FRFs are lower.

\begin{figure}[H]
	\centering
	\includegraphics[width=0.495\textwidth]{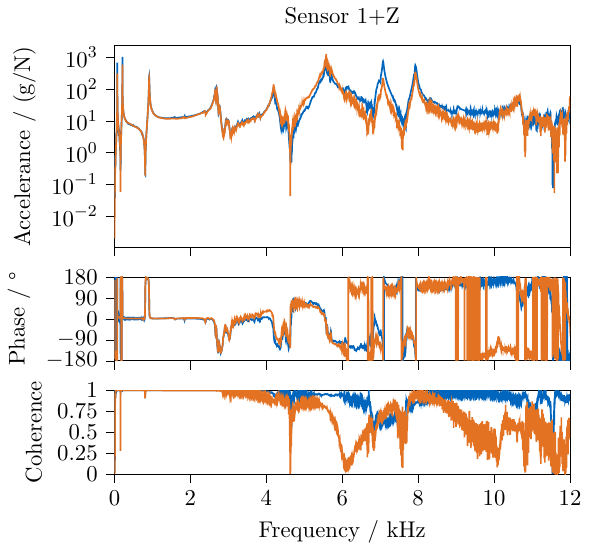}
	\includegraphics[width=0.495\textwidth]{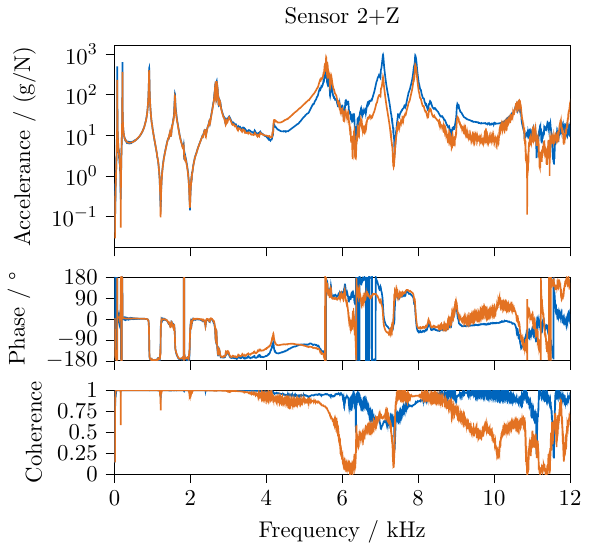} \\
	\includegraphics[width=0.495\textwidth]{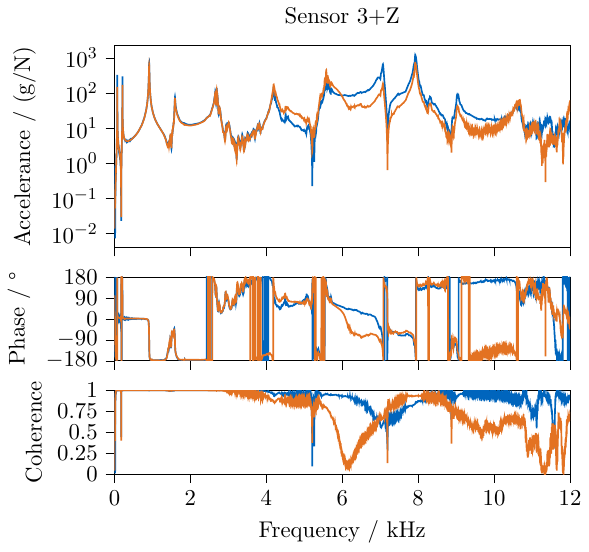}
	\includegraphics[width=0.495\textwidth]{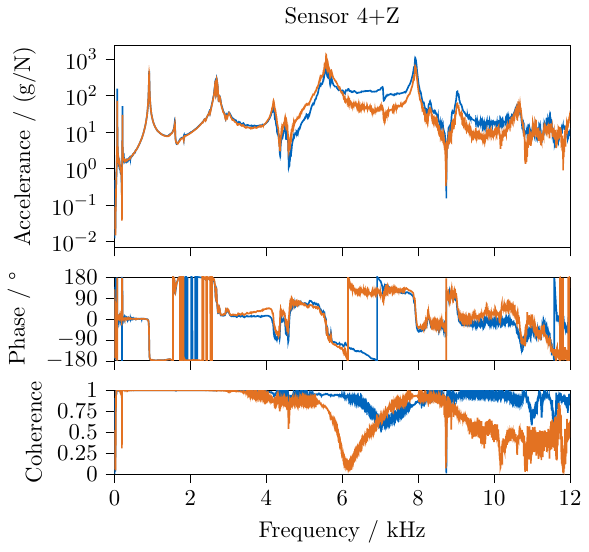}\vspace{-0.6cm}
	\caption{Comparison of FRF magnitude, phase, and coherence for sensors 1-4 between commercial and open source system, \includegraphics{figures/line_blue.pdf}$\,$\textit{Siemens LMS Scadas} System, \includegraphics{figures/line_orange.pdf}$\,$\textit{OASIS-UROS} System}
	\label{fig:FRF14}
\end{figure}

\begin{figure}[H]
	\centering
	\includegraphics[width=0.495\textwidth]{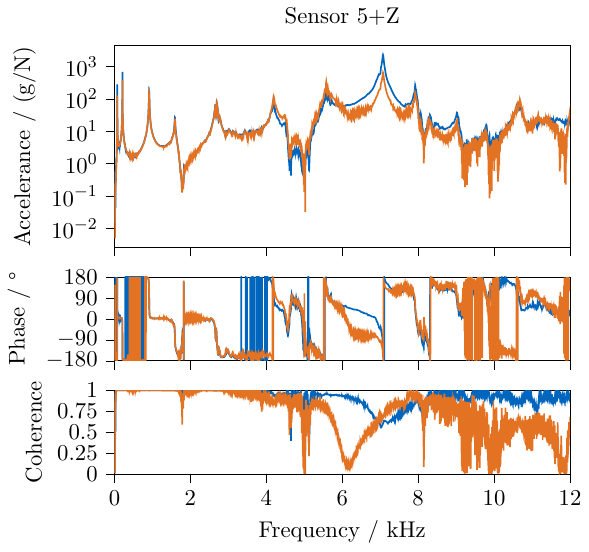}
	\includegraphics[width=0.495\textwidth]{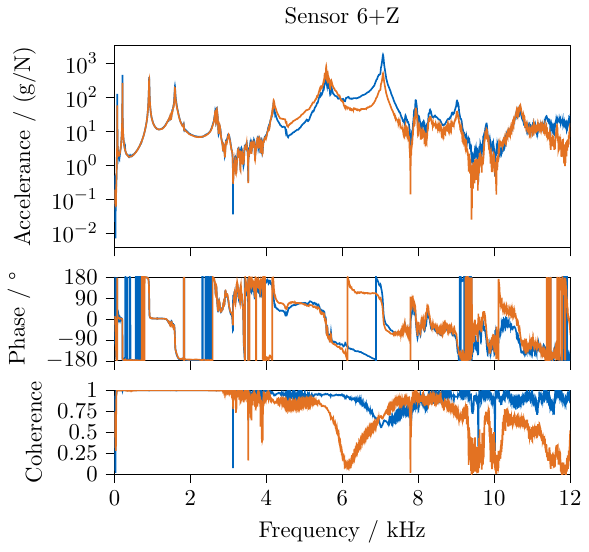} \\
	\includegraphics[width=0.495\textwidth]{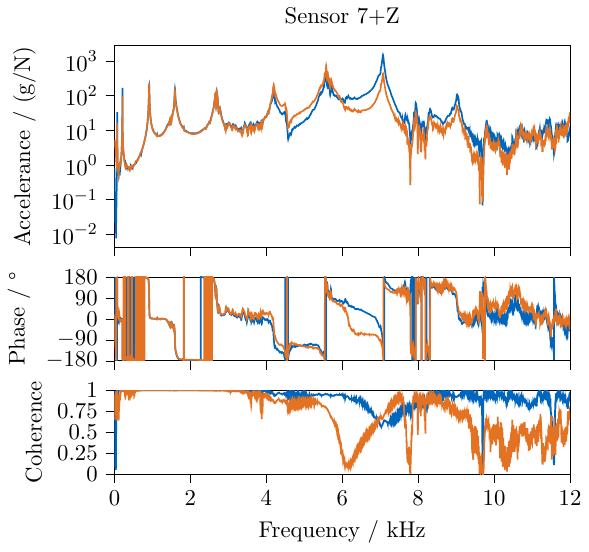}\vspace{-0.25cm}
	\caption{Comparison of FRF magnitude, phase, and coherence for sensors 5-7 between commercial and open source system, \includegraphics{figures/line_blue.pdf}$\,$\textit{Siemens LMS Scadas} System, \includegraphics{figures/line_orange.pdf}$\,$\textit{OASIS-UROS} System}
	\label{fig:FRF57}
\end{figure}

The FRFs were estimated with the $H_1$-estimator, using \textit{Simcenter Testlab} for \textit{LMS} and \textit{pyFRF}~\cite{pyFRF} for \textit{OASIS}. To compare the results, the FRF magnitude, phase, and coherence for each sensor are plotted in the following, for sensors 1 through 4 in \cref{fig:FRF14} and sensors 5 through 7 in \cref{fig:FRF57}. The results of the \textit{LMS} system are shown in blue, the ones of the \textit{OASIS} system in orange.

Considering the excitation bandwidth depicted in \cref{fig:excitation_FFT}, good results can be expected for both systems at least up to \SI{4}{\kilo\hertz}. Looking at the coherence, both systems' FRFs show basically unity coherence up to around \SI{3}{\kilo\hertz}, besides the expected drops around the anti-resonances. The magnitude and phase in this region are also very similar between the systems, besides the reduced amplitudes of \textit{OASIS} due to the response window and an increased amount of visible noise.

Starting from \SI{3}{\kilo\hertz}, a slight rise of the phase and a decrease of the coherence, not visible for the \textit{LMS} system, can be seen; for instance, this is clearly visible for sensor 1 and 4 in \cref{fig:FRF14}. At some points at the higher frequencies, e.g., between 8 and \SI{10}{\kilo\hertz} for sensor 2 (\cref{fig:FRF14}), the same phase increase can be seen. Whether this can be traced back to inaccuracies of the sampling frequency of \textit{OASIS-UROS} is currently unknown.

\vspace{-0.4cm}

For all \textit{OASIS} FRFs, a stark drop of coherence can be seen around \SI{6}{\kilo\hertz}. This could be correlated to the drop in the excitation spectrum, as shown in \cref{fig:excitation_FFT}, which is not present in the \textit{LMS} impacts. The exact reason cannot be identified here because there are too many variables involved, i.e., different impacts are used, the acquisition hardware as well as the software are different, and not all used parameters match.

In general, there are significant differences in the FRFs' magnitude and phase above around \SI{4}{\kilo\hertz}, depending on the viewed sensor. Whether this has negative effects depends on the further analysis for which the FRFs are used. For example, in the context of frequency-based substructuring, small errors can already be detrimental due to the required inversion of the FRF matrix~\cite{Allen2020}. Whether \textit{OASIS-UROS} is suitable for this application must be evaluated separately. For other analyses less sensitive to such errors, like experimental modal analysis, the observed differences in the FRFs might be less relevant. Therefore, the results of an experimental modal analysis are compared in the next section.

\vspace{-0.4cm}

\subsection{Experimental modal analysis results}

For the \textit{LMS} system, the modal identification built into \textit{Simcenter Testlab} using the \textit{PolyMAX}~\cite{PolyMAX} algorithm is used. For \textit{OASIS}, the implementation found in \textit{pyFBS}~\cite{pyFBS} is used. Here, a combination of the poly-reference Least-Squares Complex Frequency (pLSCF) and Least-Squares Frequency Domain (LSFD) methods as described in~\cite{pyFBSLSCF} is used. All modes, observable with the test setup, until \SI{8}{\kilo\hertz} are identified. For this, two modal fits are performed, once from 1 to \SI{4000}{\hertz} and once from 4 to \SI{8.3}{\kilo\hertz}.

The extracted eigenfrequencies and damping ratios are summarized in \cref{tab:EMA}. Additionally, the difference between the \textit{OASIS} results and the \textit{LMS} results is given. For the eigenfrequencies, the relative difference in percent is given, while for the damping ratios, the absolute difference of the damping ratio is given. Note, that this does not compare the two modal fit algorithms, but the entire acquisition and post-processing stack.

\begin{table}[H]
	\centering
	\begin{tabular}{c|c|c|c|c|c|c}
		 & \multicolumn{3}{c|}{\cellcolor{TUMBlue}\textbf{\textcolor{white}{Eigenfrequencies $f_i$}}} & \multicolumn{3}{c}{\cellcolor{TUMBlue}\textbf{\textcolor{white}{Damping ratios $\vartheta_i$}}} \\\cline{2-7}
		 & LMS & OASIS & Relative Difference & LMS & OASIS & Absolute Difference \\\hline
		 Mode 1 & \SI{74.8}{\hertz} & \SI{74.8}{\hertz} & \SI{0.00}{\percent} & \SI{0.16}{\percent} & \SI{0.13}{\percent} & +\SI{0.03}{\percent} \\\hline
		 Mode 2 & \SI{209.1}{\hertz} & \SI{209.2}{\hertz} & \SI{0.00}{\percent} & \SI{0.14}{\percent} & \SI{0.14}{\percent} & \SI{0.00}{\percent} \\\hline
		 Mode 3 & \SI{914.5}{\hertz} & \SI{913.4}{\hertz} & -\SI{0.12}{\percent} & \SI{0.44}{\percent} & \SI{0.42}{\percent} & -\SI{0.02}{\percent} \\\hline
		 Mode 4 & \SI{1594.0}{\hertz} & \SI{1593.4}{\hertz} & -\SI{0.04}{\percent} & \SI{0.31}{\percent} & \SI{0.73}{\percent} & +\SI{0.42}{\percent} \\\hline
		 Mode 5 & \SI{2662.3}{\hertz} & \SI{2669.5}{\hertz} & +\SI{0.27}{\percent} & \SI{0.52}{\percent} & \SI{0.46}{\percent} & -\SI{0.06}{\percent} \\\hline
		 Mode 6 & \SI{2692.5}{\hertz} & \SI{2687.1}{\hertz} & -\SI{0.20}{\percent} & \SI{0.51}{\percent} & \SI{0.43}{\percent} & -\SI{0.08}{\percent} \\\hline
		 Mode 7 & \SI{4189.7}{\hertz} & \SI{4198.4}{\hertz} & +\SI{0.20}{\percent} & \SI{1.03}{\percent} & \SI{0.72}{\percent} & -\SI{0.31}{\percent} \\\hline
		 Mode 8 & \SI{5576.8}{\hertz} & \SI{5578.9}{\hertz} & +\SI{0.04}{\percent} & \SI{0.55}{\percent} & \SI{1.06}{\percent} & +\SI{0.51}{\percent} \\\hline
		 Mode 9 & \SI{7076.7}{\hertz} & \SI{7075.3}{\hertz} & -\SI{0.02}{\percent} & \SI{0.36}{\percent} & \SI{0.39}{\percent} & +\SI{0.03}{\percent} \\\hline
		 Mode 10 & \SI{7930.8}{\hertz} & \SI{7931.8}{\hertz} & +\SI{0.01}{\percent} & \SI{0.24}{\percent} & \SI{0.27}{\percent} & +\SI{0.03}{\percent}
	\end{tabular}
	\caption{Summary of experimental modal analysis comparison of commercial system (\textit{Siemens LMS Scadas System} and \textit{PolyMAX}~\cite{PolyMAX}) with open source system (\textit{OASIS-UROS} and \textit{pyFBS} modal identification~\cite{pyFBSLSCF})}
	\label{tab:EMA}
\end{table}\vspace{-0.4cm}

For the eigenfrequencies, the most significant deviations can be seen for modes 5 to 7, where the highest deviation occurs for mode 5 with +\SI{0.27}{\percent} or +\SI{7.2}{\hertz}. The higher modes, i.e., modes 8 through 10, show less discrepancy, both, in relative and absolute values. This indicates that there is no significant shift with increasing frequency, as could result from an inaccurate sampling clock of the \textit{OASIS} system.

For the damping ratios, greater differences can be observed, especially for modes 4, 7, and 8. As previously mentioned, the exact reason for this cannot be pinpointed here due to the entirely  different acquisition and post-processing setups. A separate analysis, where, for example, only the acquisition hardware differs, would have to be performed. However, the comparison performed here evaluates the real-world use case, where only free software would be used in conjunction with \textit{OASIS}.

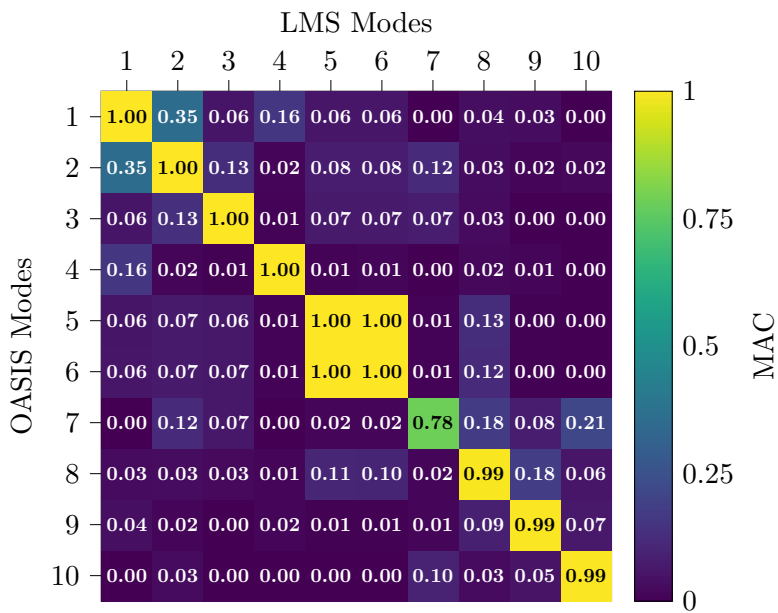
\begin{figure}[H]
	\centering
	% This file was created with tikzplotlib v0.10.1.
\begin{tikzpicture}

\definecolor{darkgray176}{RGB}{176,176,176}

\begin{axis}[
colorbar,
colorbar style={ylabel={MAC},ytick={0,0.25,0.5,0.75,1},ymin=0, ymax=1,major tick length=0.0cm},
colormap/viridis,
height=0.45\textwidth,
point meta max=0.999958880457599,
point meta min=2.99752354059061e-05,
tick align=outside,
width=0.45\textwidth,
x grid style={darkgray176},
xlabel={LMS Modes},
xmin=0.5, xmax=10.5,
xtick pos=right,
xtick={1,2,3,4,5,6,7,8,9,10},
xtick style={color=black},
y dir=reverse,
y grid style={darkgray176},
ylabel={OASIS Modes},
ymin=0.5, ymax=10.5,
ytick pos=left,
ytick={1,2,3,4,5,6,7,8,9,10},
ytick style={color=black}
]
\addplot graphics [includegraphics cmd=\pgfimage,xmin=0.5, xmax=10.5, ymin=10.5, ymax=0.5] {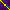};
\draw (axis cs:1,1) node[
  scale=0.69,
  text=black,
  rotate=0.0
]{\bfseries 1.00};
\draw (axis cs:1,2) node[
  scale=0.69,
  text=white,
  rotate=0.0
]{\bfseries 0.35};
\draw (axis cs:1,3) node[
  scale=0.69,
  text=white,
  rotate=0.0
]{\bfseries 0.06};
\draw (axis cs:1,4) node[
  scale=0.69,
  text=white,
  rotate=0.0
]{\bfseries 0.16};
\draw (axis cs:1,5) node[
  scale=0.69,
  text=white,
  rotate=0.0
]{\bfseries 0.06};
\draw (axis cs:1,6) node[
  scale=0.69,
  text=white,
  rotate=0.0
]{\bfseries 0.06};
\draw (axis cs:1,7) node[
  scale=0.69,
  text=white,
  rotate=0.0
]{\bfseries 0.00};
\draw (axis cs:1,8) node[
  scale=0.69,
  text=white,
  rotate=0.0
]{\bfseries 0.03};
\draw (axis cs:1,9) node[
  scale=0.69,
  text=white,
  rotate=0.0
]{\bfseries 0.04};
\draw (axis cs:1,10) node[
  scale=0.69,
  text=white,
  rotate=0.0
]{\bfseries 0.00};
\draw (axis cs:2,1) node[
  scale=0.69,
  text=white,
  rotate=0.0
]{\bfseries 0.35};
\draw (axis cs:2,2) node[
  scale=0.69,
  text=black,
  rotate=0.0
]{\bfseries 1.00};
\draw (axis cs:2,3) node[
  scale=0.69,
  text=white,
  rotate=0.0
]{\bfseries 0.13};
\draw (axis cs:2,4) node[
  scale=0.69,
  text=white,
  rotate=0.0
]{\bfseries 0.02};
\draw (axis cs:2,5) node[
  scale=0.69,
  text=white,
  rotate=0.0
]{\bfseries 0.07};
\draw (axis cs:2,6) node[
  scale=0.69,
  text=white,
  rotate=0.0
]{\bfseries 0.07};
\draw (axis cs:2,7) node[
  scale=0.69,
  text=white,
  rotate=0.0
]{\bfseries 0.12};
\draw (axis cs:2,8) node[
  scale=0.69,
  text=white,
  rotate=0.0
]{\bfseries 0.03};
\draw (axis cs:2,9) node[
  scale=0.69,
  text=white,
  rotate=0.0
]{\bfseries 0.02};
\draw (axis cs:2,10) node[
  scale=0.69,
  text=white,
  rotate=0.0
]{\bfseries 0.03};
\draw (axis cs:3,1) node[
  scale=0.69,
  text=white,
  rotate=0.0
]{\bfseries 0.06};
\draw (axis cs:3,2) node[
  scale=0.69,
  text=white,
  rotate=0.0
]{\bfseries 0.13};
\draw (axis cs:3,3) node[
  scale=0.69,
  text=black,
  rotate=0.0
]{\bfseries 1.00};
\draw (axis cs:3,4) node[
  scale=0.69,
  text=white,
  rotate=0.0
]{\bfseries 0.01};
\draw (axis cs:3,5) node[
  scale=0.69,
  text=white,
  rotate=0.0
]{\bfseries 0.06};
\draw (axis cs:3,6) node[
  scale=0.69,
  text=white,
  rotate=0.0
]{\bfseries 0.07};
\draw (axis cs:3,7) node[
  scale=0.69,
  text=white,
  rotate=0.0
]{\bfseries 0.07};
\draw (axis cs:3,8) node[
  scale=0.69,
  text=white,
  rotate=0.0
]{\bfseries 0.03};
\draw (axis cs:3,9) node[
  scale=0.69,
  text=white,
  rotate=0.0
]{\bfseries 0.00};
\draw (axis cs:3,10) node[
  scale=0.69,
  text=white,
  rotate=0.0
]{\bfseries 0.00};
\draw (axis cs:4,1) node[
  scale=0.69,
  text=white,
  rotate=0.0
]{\bfseries 0.16};
\draw (axis cs:4,2) node[
  scale=0.69,
  text=white,
  rotate=0.0
]{\bfseries 0.02};
\draw (axis cs:4,3) node[
  scale=0.69,
  text=white,
  rotate=0.0
]{\bfseries 0.01};
\draw (axis cs:4,4) node[
  scale=0.69,
  text=black,
  rotate=0.0
]{\bfseries 1.00};
\draw (axis cs:4,5) node[
  scale=0.69,
  text=white,
  rotate=0.0
]{\bfseries 0.01};
\draw (axis cs:4,6) node[
  scale=0.69,
  text=white,
  rotate=0.0
]{\bfseries 0.01};
\draw (axis cs:4,7) node[
  scale=0.69,
  text=white,
  rotate=0.0
]{\bfseries 0.00};
\draw (axis cs:4,8) node[
  scale=0.69,
  text=white,
  rotate=0.0
]{\bfseries 0.01};
\draw (axis cs:4,9) node[
  scale=0.69,
  text=white,
  rotate=0.0
]{\bfseries 0.02};
\draw (axis cs:4,10) node[
  scale=0.69,
  text=white,
  rotate=0.0
]{\bfseries 0.00};
\draw (axis cs:5,1) node[
  scale=0.69,
  text=white,
  rotate=0.0
]{\bfseries 0.06};
\draw (axis cs:5,2) node[
  scale=0.69,
  text=white,
  rotate=0.0
]{\bfseries 0.08};
\draw (axis cs:5,3) node[
  scale=0.69,
  text=white,
  rotate=0.0
]{\bfseries 0.07};
\draw (axis cs:5,4) node[
  scale=0.69,
  text=white,
  rotate=0.0
]{\bfseries 0.01};
\draw (axis cs:5,5) node[
  scale=0.69,
  text=black,
  rotate=0.0
]{\bfseries 1.00};
\draw (axis cs:5,6) node[
  scale=0.69,
  text=black,
  rotate=0.0
]{\bfseries 1.00};
\draw (axis cs:5,7) node[
  scale=0.69,
  text=white,
  rotate=0.0
]{\bfseries 0.02};
\draw (axis cs:5,8) node[
  scale=0.69,
  text=white,
  rotate=0.0
]{\bfseries 0.11};
\draw (axis cs:5,9) node[
  scale=0.69,
  text=white,
  rotate=0.0
]{\bfseries 0.01};
\draw (axis cs:5,10) node[
  scale=0.69,
  text=white,
  rotate=0.0
]{\bfseries 0.00};
\draw (axis cs:6,1) node[
  scale=0.69,
  text=white,
  rotate=0.0
]{\bfseries 0.06};
\draw (axis cs:6,2) node[
  scale=0.69,
  text=white,
  rotate=0.0
]{\bfseries 0.08};
\draw (axis cs:6,3) node[
  scale=0.69,
  text=white,
  rotate=0.0
]{\bfseries 0.07};
\draw (axis cs:6,4) node[
  scale=0.69,
  text=white,
  rotate=0.0
]{\bfseries 0.01};
\draw (axis cs:6,5) node[
  scale=0.69,
  text=black,
  rotate=0.0
]{\bfseries 1.00};
\draw (axis cs:6,6) node[
  scale=0.69,
  text=black,
  rotate=0.0
]{\bfseries 1.00};
\draw (axis cs:6,7) node[
  scale=0.69,
  text=white,
  rotate=0.0
]{\bfseries 0.02};
\draw (axis cs:6,8) node[
  scale=0.69,
  text=white,
  rotate=0.0
]{\bfseries 0.10};
\draw (axis cs:6,9) node[
  scale=0.69,
  text=white,
  rotate=0.0
]{\bfseries 0.01};
\draw (axis cs:6,10) node[
  scale=0.69,
  text=white,
  rotate=0.0
]{\bfseries 0.00};
\draw (axis cs:7,1) node[
  scale=0.69,
  text=white,
  rotate=0.0
]{\bfseries 0.00};
\draw (axis cs:7,2) node[
  scale=0.69,
  text=white,
  rotate=0.0
]{\bfseries 0.12};
\draw (axis cs:7,3) node[
  scale=0.69,
  text=white,
  rotate=0.0
]{\bfseries 0.07};
\draw (axis cs:7,4) node[
  scale=0.69,
  text=white,
  rotate=0.0
]{\bfseries 0.00};
\draw (axis cs:7,5) node[
  scale=0.69,
  text=white,
  rotate=0.0
]{\bfseries 0.01};
\draw (axis cs:7,6) node[
  scale=0.69,
  text=white,
  rotate=0.0
]{\bfseries 0.01};
\draw (axis cs:7,7) node[
  scale=0.69,
  text=black,
  rotate=0.0
]{\bfseries 0.78};
\draw (axis cs:7,8) node[
  scale=0.69,
  text=white,
  rotate=0.0
]{\bfseries 0.02};
\draw (axis cs:7,9) node[
  scale=0.69,
  text=white,
  rotate=0.0
]{\bfseries 0.01};
\draw (axis cs:7,10) node[
  scale=0.69,
  text=white,
  rotate=0.0
]{\bfseries 0.10};
\draw (axis cs:8,1) node[
  scale=0.69,
  text=white,
  rotate=0.0
]{\bfseries 0.04};
\draw (axis cs:8,2) node[
  scale=0.69,
  text=white,
  rotate=0.0
]{\bfseries 0.03};
\draw (axis cs:8,3) node[
  scale=0.69,
  text=white,
  rotate=0.0
]{\bfseries 0.03};
\draw (axis cs:8,4) node[
  scale=0.69,
  text=white,
  rotate=0.0
]{\bfseries 0.02};
\draw (axis cs:8,5) node[
  scale=0.69,
  text=white,
  rotate=0.0
]{\bfseries 0.13};
\draw (axis cs:8,6) node[
  scale=0.69,
  text=white,
  rotate=0.0
]{\bfseries 0.12};
\draw (axis cs:8,7) node[
  scale=0.69,
  text=white,
  rotate=0.0
]{\bfseries 0.18};
\draw (axis cs:8,8) node[
  scale=0.69,
  text=black,
  rotate=0.0
]{\bfseries 0.99};
\draw (axis cs:8,9) node[
  scale=0.69,
  text=white,
  rotate=0.0
]{\bfseries 0.09};
\draw (axis cs:8,10) node[
  scale=0.69,
  text=white,
  rotate=0.0
]{\bfseries 0.03};
\draw (axis cs:9,1) node[
  scale=0.69,
  text=white,
  rotate=0.0
]{\bfseries 0.03};
\draw (axis cs:9,2) node[
  scale=0.69,
  text=white,
  rotate=0.0
]{\bfseries 0.02};
\draw (axis cs:9,3) node[
  scale=0.69,
  text=white,
  rotate=0.0
]{\bfseries 0.00};
\draw (axis cs:9,4) node[
  scale=0.69,
  text=white,
  rotate=0.0
]{\bfseries 0.01};
\draw (axis cs:9,5) node[
  scale=0.69,
  text=white,
  rotate=0.0
]{\bfseries 0.00};
\draw (axis cs:9,6) node[
  scale=0.69,
  text=white,
  rotate=0.0
]{\bfseries 0.00};
\draw (axis cs:9,7) node[
  scale=0.69,
  text=white,
  rotate=0.0
]{\bfseries 0.08};
\draw (axis cs:9,8) node[
  scale=0.69,
  text=white,
  rotate=0.0
]{\bfseries 0.18};
\draw (axis cs:9,9) node[
  scale=0.69,
  text=black,
  rotate=0.0
]{\bfseries 0.99};
\draw (axis cs:9,10) node[
  scale=0.69,
  text=white,
  rotate=0.0
]{\bfseries 0.05};
\draw (axis cs:10,1) node[
  scale=0.69,
  text=white,
  rotate=0.0
]{\bfseries 0.00};
\draw (axis cs:10,2) node[
  scale=0.69,
  text=white,
  rotate=0.0
]{\bfseries 0.02};
\draw (axis cs:10,3) node[
  scale=0.69,
  text=white,
  rotate=0.0
]{\bfseries 0.00};
\draw (axis cs:10,4) node[
  scale=0.69,
  text=white,
  rotate=0.0
]{\bfseries 0.00};
\draw (axis cs:10,5) node[
  scale=0.69,
  text=white,
  rotate=0.0
]{\bfseries 0.00};
\draw (axis cs:10,6) node[
  scale=0.69,
  text=white,
  rotate=0.0
]{\bfseries 0.00};
\draw (axis cs:10,7) node[
  scale=0.69,
  text=white,
  rotate=0.0
]{\bfseries 0.21};
\draw (axis cs:10,8) node[
  scale=0.69,
  text=white,
  rotate=0.0
]{\bfseries 0.06};
\draw (axis cs:10,9) node[
  scale=0.69,
  text=white,
  rotate=0.0
]{\bfseries 0.07};
\draw (axis cs:10,10) node[
  scale=0.69,
  text=black,
  rotate=0.0
]{\bfseries 0.99};
\end{axis}

\end{tikzpicture}
	\caption{Cross-Modal Assurance Criterion between the mode shapes identified by the commercial system (\textit{Siemens LMS Scadas System}) and the open source system (\textit{OASIS-UROS})}
	\label{fig:MAC}
\end{figure}

Lastly, the estimated mode shapes are compared by building the Cross-Modal Assurance Criterion (Cross-MAC) between the two mode sets. The resulting MAC matrix is visualized in \cref{fig:MAC}. As can be seen from the close to unity values on the diagonal, both mode shape sets match pretty closely. Only mode 7 shows a lower MAC value, which might be resolvable by selecting another pole from the stabilization diagram.

\vspace{-0.4cm}

\newpage

To summarize, for frequencies up to \SI{3}{\kilo\hertz}, no significant differences were observable in the FRFs estimated by the commercial system with proprietary software and those retrieved using open source software and hardware. Besides greater differences in the damping ratios, the results of an experimental modal analysis matched also pretty closely between commercial and open source system. While \textit{OASIS-UROS} cannot match the performance of the commercial system, the authors believe that, especially when comparing the costs, the developed system is a viable alternative for students, people in academia, or smaller companies that have a constrained budget or require complete insight as well as adaptability of the hardware and software.

% !TeX spellcheck = en_US

\noindent
\textbf{CRediT author statement}\\
\noindent
\textbf{O.~M.~Zobel:} Conceptualization; Methodology; Software; Validation; Formal Analysis; Investigation; Data Curation; Writing - Original Draft; Writing - Review \& Editing; Visualization;
\textbf{J.~Maierhofer:} Conceptualization; Methodology; Validation; Formal Analysis; Investigation; Writing - Original Draft; Writing - Review \& Editing.
\textbf{A.~Köstler:} Validation; Investigation; Resources; Writing - Review \& Editing. 
\textbf{D.~J.~Rixen:} Resources, Writing - review \& editing, Supervision.\\
% !TeX spellcheck = en_US

\noindent
\textbf{Acknowledgements}\\
This research did not receive any specific grant from funding agencies in the public, commercial, or not-for-profit sectors.\\

\noindent\textbf{References}
\printbibliography[heading=none]

\newpage
\appendix
% !TeX spellcheck = en_US

\section{Command reference}\label{sec:CommandRef}

Note: Included here is the command reference at the point of publication, the current version can be found at: \url{https://gitlab.com/oasis-acquisition/oasis-commandreference}

\begin{center}
	\vfill
	\fbox{\includegraphics[scale=0.72,page=1]{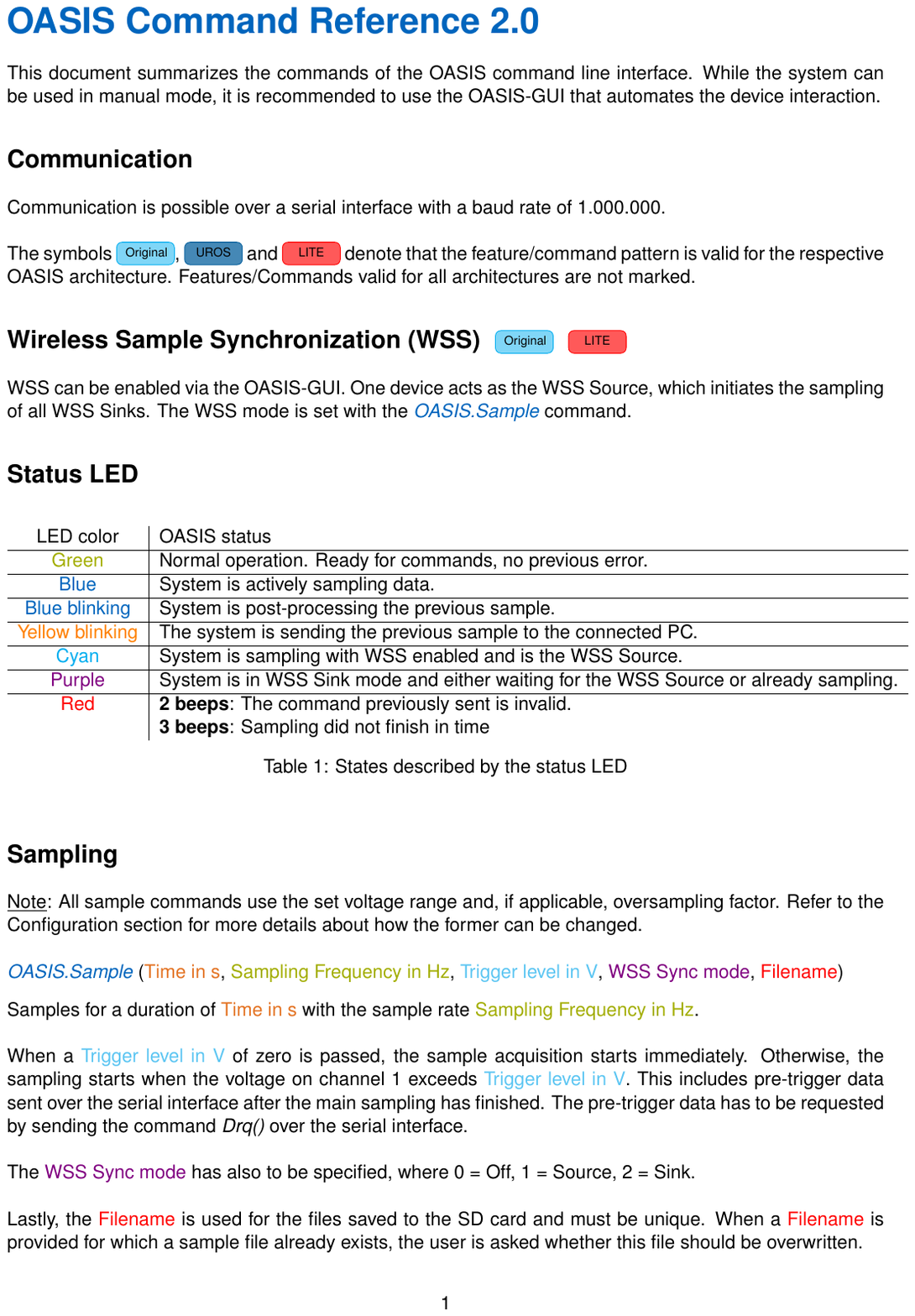}}
	\vfill
\end{center}
\newpage
$ $
\begin{center}
	\vfill
	\fbox{\includegraphics[scale=0.72,page=2]{OASIS-Command-Reference.pdf}}
	\vfill
\end{center}
\newpage
$ $
\begin{center}
	\vfill
	\fbox{\includegraphics[scale=0.72,page=3]{OASIS-Command-Reference.pdf}}
	\vfill
\end{center}
\newpage

\end{document}